\documentclass[twocolumn,showpacs]{revtex4} %twocolumn
\usepackage{graphicx}
\usepackage{setspace} % for the \setstretch{2} command
\usepackage{bm} % bold math
\usepackage{amssymb} % use this package to enable \nrightarrow command
\usepackage{amsmath} % use this packale to enable \xrightarrow command

\begin{document}
%\setstretch{2}

\title{Shockley model description of surface states in topological insulators}

\author{Sergey S. Pershoguba and  Victor M. Yakovenko}

\affiliation{Center for Nanophysics and Advanced Materials, Department of Physics, University of Maryland, College Park, Maryland 20742-4111, USA}

%\date{v. U , edited by VMY on July 22, 2012, compiled \today}

\date{\today}

\begin{abstract}
Surface states in topological insulators can be understood based on the well-known Shockley model, a one-dimensional tight-binding model with two atoms per elementary cell, connected via alternating tunneling amplitudes. We  generalize the one-dimensional model to the three-dimensional case representing a sequence of layers connected via tunneling amplitudes $t$, which depend on the in-plane momentum $\bm p = (p_x,p_y)$. The Hamiltonian of the model is a $2\times 2$ matrix with the  off-diagonal element $t(k,\bm p)$ depending also on the out-of-plane momentum $k$. We show that the existence of the surface states depends on the complex function $t(k,\bm p)$. The surface states exist for those in-plane momenta $\bm p$ where the winding  number of the function $t(k,\bm p)$ is non-zero when $k$ is changed from $0$ to $2\pi$. The sign of the winding number determines the sublattice on which the surface states are localized. The equation $t(k,\bm p)=0$ defines a vortex line in the three-dimensional momentum space. Projection of the vortex line onto the space of the two-dimensional momentum $\bm p$ encircles the domain where the surface states exist. We illustrate how this approach works for a well-known model of a topological insulator on the diamond lattice. We find that different configurations of the vortex lines are responsible for the ``weak'' and ``strong'' topological insulator  phases. A topological transition occurs when the vortex lines reconnect from spiral to circular form. We apply the Shockley model to Bi$_2$Se$_3$ and discuss applicability of a continuous approximation for the description of the surface states. We conclude that the tight-binding model gives a better description of the surface states.
\end{abstract}


\maketitle

%%%%%%%%%%%%%%%%%%%%%%%%%%%%%%%%%%%%%%%%%%%%%%%%%%%%%%%%%%%%%%%%%%%%%%%%%%%%%
\section{Introduction} \label{sec:intro}
%%%%%%%%%%%%%%%%%%%%%%%%%%%%%%%%%%%%%%%%%%%%%%%%%%%%%%%%%%%%%%%%%%%%%%%%%%%%%
%%%%%%%%%%%%%%%%%%%%%%%%%%%%%%%%%%%%%%%%%%%%%%%%%%%%%%%%%%%%%%%%%%%%%%%%%%%%%
\begin{figure*}
\centering
\hbox{
\includegraphics[height=0.3\linewidth]{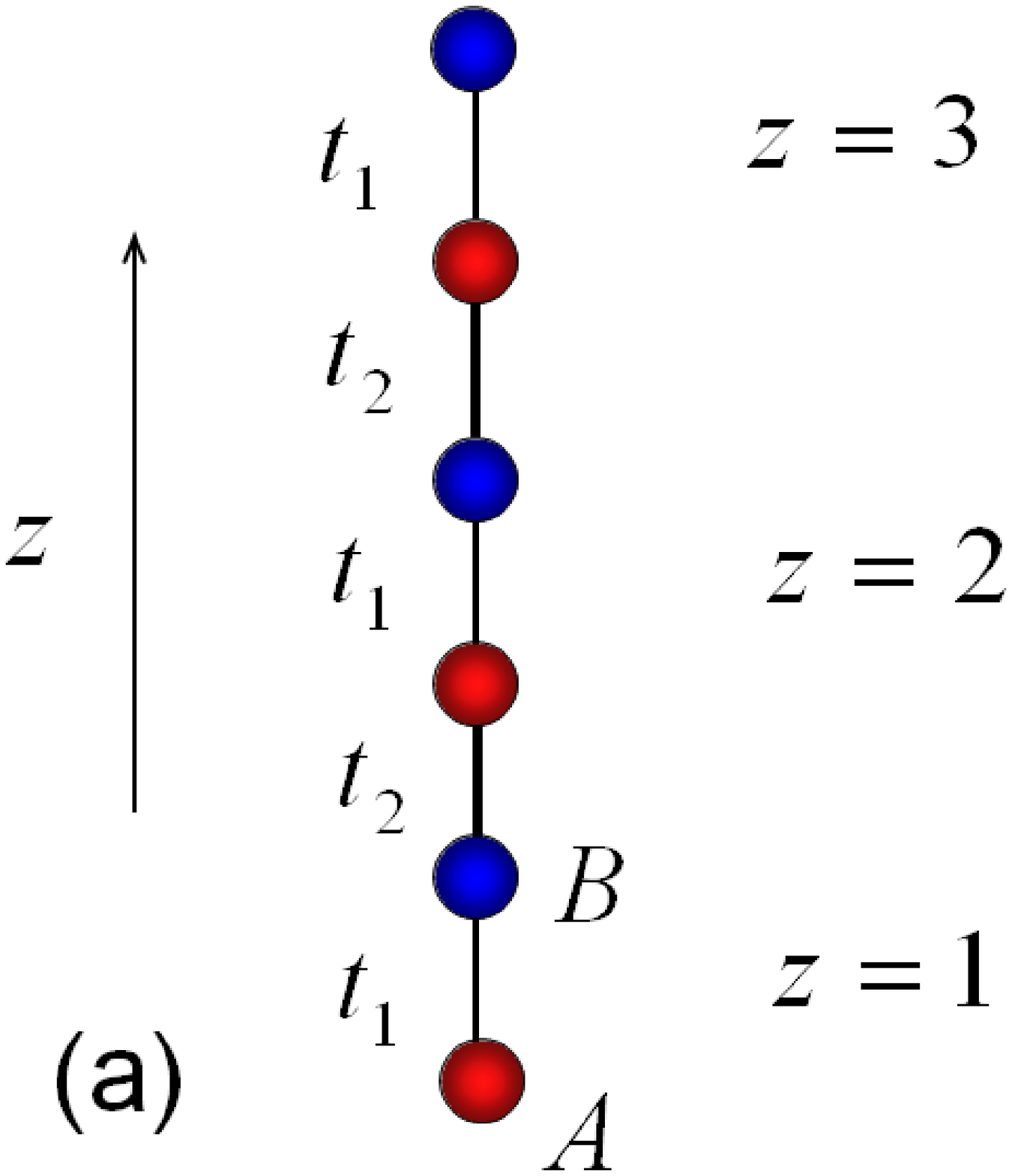} \hspace{0in}
\includegraphics[height=0.3\linewidth]{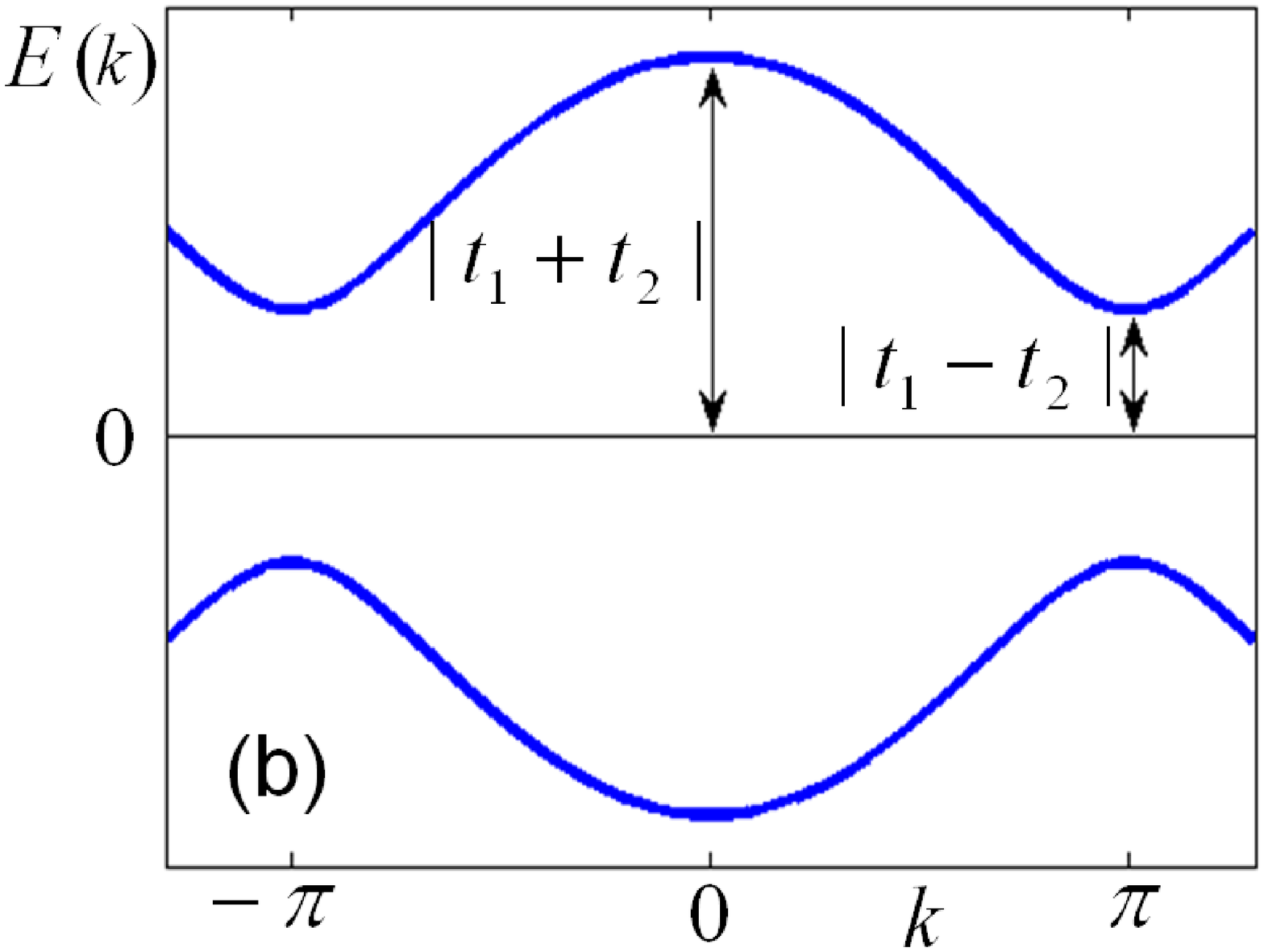} \hspace{0in}
\includegraphics[height=0.3\linewidth]{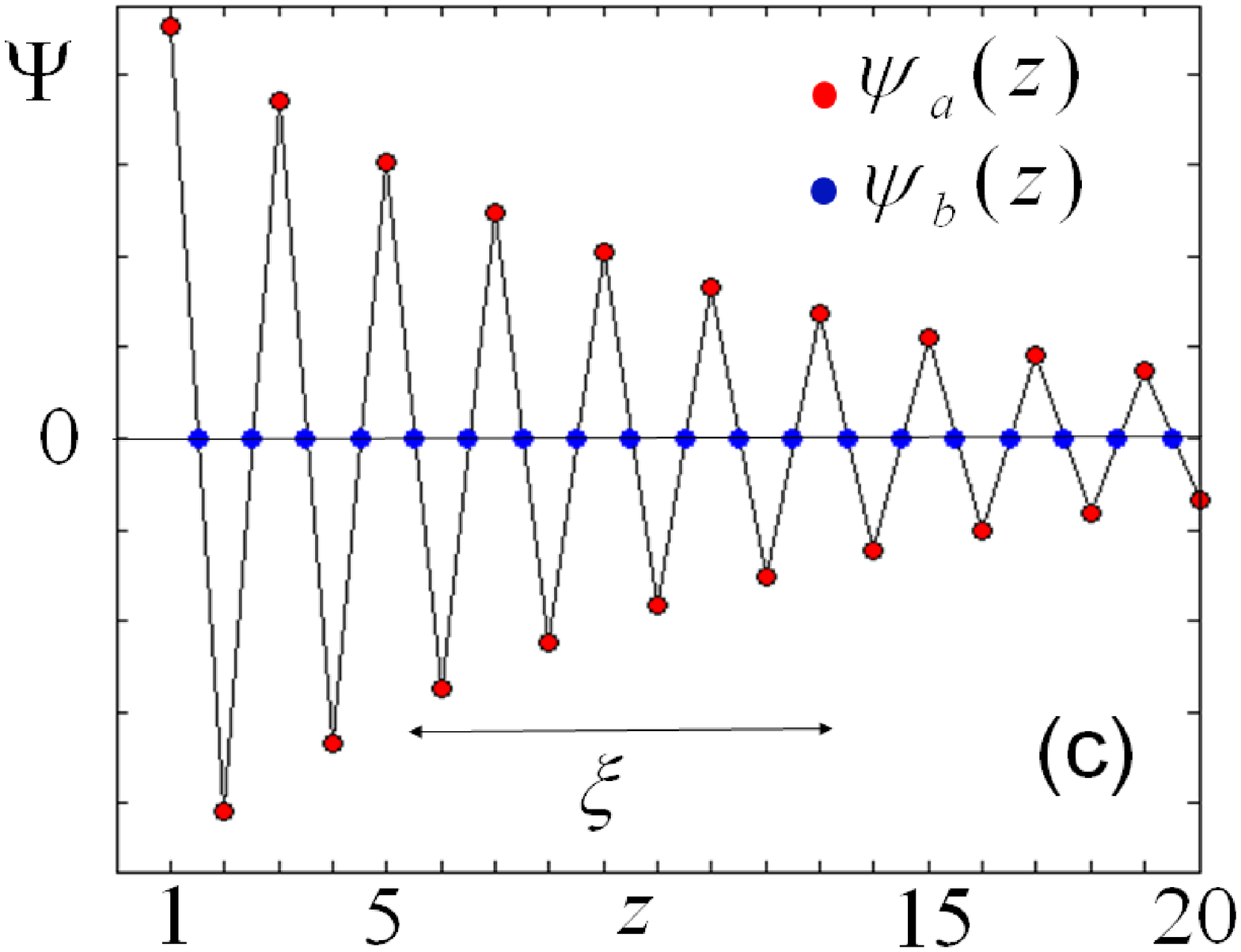} \hspace{0in}
}
\caption{(Color online) Panel (a): 1D chain of atoms with alternating tunneling amplitudes $t_1$ and $t_2$ representing the Shockley model, Eq.~(\ref{Ham}). Panel (b): The bulk energy spectrum of the system, Eq.~(\ref{spectr}), with a non-zero gap for $|t_1|\neq |t_2|$. Panel (c): The exponentially decaying edge state, Eq.~(\ref{psi}),  for $|t_1|/|t_2|<1$ with the penetration depth $\xi=1/{\rm ln}|t_2/t_1|$.} \label{fig:Shockley}
\end{figure*}
%%%%%%%%%%%%%%%%%%%%%%%%%%%%%%%%%%%%%%%%%%%%%%%%%%%%%%%%%%%%%%%%%%%%%%%%%%%%%

Recent theoretical discovery \cite{Mele-2005,Kane-2005,Kane-2007,Fu-2007,Bernevig-2006,Bernevig-2006a,Qi-2008,Moore-2007,Roy-2009,Roy-2009a,Schnyder-2008}  of topological insulators has stimulated active research in the field \cite{Hasan-2010,Hasan-2010a,Qi-2011}. The key idea is that a time-reversal-invariant band Hamiltonian with a finite gap over the Brillouin zone (BZ) can be characterized by the topological indices $Z_2$ \cite{Kane-2005,Kane-2007,Fu-2007}. The topological indices distinguish trivial and non-trivial phases, which usually arise due to strong spin-orbit coupling in the system \cite{Mele-2005,Bernevig-2006a}. The topological $Z_2$ indices are robust to moderate perturbations of the Hamiltonian and can change only if the energy gap is closed \cite{Kane-2005,Kane-2007,Fu-2007}. Since the topological indices of vacuum and TI are different \cite{Hasan-2010,Volovik-2003}, the boundary of TI should carry gapless modes \cite{Volkov-1985,Pankratov-1987}, similar to the chiral edge modes in the quantum Hall phases \cite{Haldane-1987}. Because of the bulk-boundary correspondence, the gapless surface states are topologically protected against moderate perturbations.

Theory of topological surface states has been studied in a number of works both in the tight-binding \cite{Hosur-2010,Mong-2011} and continuous models \cite{H-Zhang-2009,Liu-2010,Linder-2009,Shen-2010,Shan-2010a,Lu-2010}. Many papers focused on the bulk-boundary correspondence, i.e., on proving that a sample with non-trivial topological numbers in the bulk should possess gapless excitations on the surface.  The method of topological invariants, although being very powerful, is often not physically transparent and not intuitive about the exact mechanism by which the topological numbers are related to the surface states.

The purpose of our paper is to show that formation of the surface states in TIs can be understood based on the simple and well-known Shockley model \cite{Shockley-1939,Davison-1996} of the edge states. The Shockley model was also applied to surface states in topological superconductors \cite{Wimmer-2010,Beenakker-2011}; however, we focus only on surfaces states in semiconductors. In Sec.~\ref{sec:Shockley}, we review the one-dimensional (1D) Shockley model consisting of a chain of  atoms connected via alternating tight-binding hopping amplitudes $t_1$ and $t_2$. When a boundary is introduced in the system, e.g. by breaking the bond $t_2$, existence of the edge states is governed by the Shockley criterion. The edge state exists if the greater tight-binding amplitude is broken at the boundary, i.e. if $|t_2|>|t_1|$, and it is localized on one sublattice. In the end of Sec.~\ref{sec:classicalSchockley}, we show how the Shockley criterion can be formulated in terms of a topological winding number for the off-diagonal matrix element of the bulk Hamiltonian, thus connecting bulk properties with the surface states as discussed in Refs.~\cite{Mong-2011,Heikkila-2011,Ryu-2002,Deplace-2011,Gurarie-2011}. In Sec.~\ref{sec:1Dto3D}, we generalize the model to three dimensions (3D) by replacing atoms by the two-dimensional (2D) layers parallel to the $xy$ plane and assigning the in-plane momentum  dependence $\bm p=(p_x,p_y)$ to the interlayer tunneling amplitudes $t_1$ and $t_2$. In Sec.~\ref{sec:VortexLines}, we study vortex lines in the 3D momentum space~\cite{Beri-2010,Schnyder-2010}, where the off-diagonal matrix element of the bulk Hamiltonian vanishes. We show that the projection of the 3D vortex lines  onto the 2D in-plane momentum space encircles the domain where the surface states exist. We observe that the tight-binding TI Hamiltonians studied in Refs. \cite{Kane-2005,Bernevig-2006,Kane-2007,Hosur-2010} have the Shockley-model structure and can be understood using our approach. In Sec.~\ref{sec:DiamondHamiltonian}, we illustrate the Shockley mechanism for the Fu-Kane-Mele model on the diamond lattice \cite{Kane-2007}. We show how the surface states evolve when the parameters of the Hamiltonian vary. In Sec.~\ref{sec:Vortex}, we show that reconnection of the vortex lines represents a phase transition in the TI Hamiltonian. The spiral vortex lines correspond to a phase with an even number of Dirac cones (the ``weak'' TI phase), while the circular vortex lines correspond to a phase with an odd number of Dirac cones (the ``strong'' TI phase). In Sec.~\ref{sec:BiSeShockley}, we apply the Shockley model to describe the surface states in Bi$_2$Se$_3$, which is formed by the quintuple layers of Bi and Se \cite{H-Zhang-2009,Liu-2010,Fu-2011,DHsieh-2009,Analytis-2010,Hor-2010}. The electronic structure of this material near the Fermi level can be well described by the hybridized $p_z$ orbitals located near the outer layers of the quintuplets \cite{Liu-2010,Fu-2011}. Thus, the Shockley model with the intra-quintuplet and inter-quintuplet tunneling amplitudes $t_1$ ant $t_2$ gives a plausible description of this material. Surface states have complementary properties depending on how the crystal is terminated~\cite{Fu-2011}. Breaking the $t_2$ amplitude introduces a cut between the quintuplets. In this case, the surface states have a Dirac cone in the Brillouin zone (BZ) center~\cite{DHsieh-2009,Analytis-2010}. Breaking $t_1$ introduces a cut inside the quintuplet. In this case, the Shockley model predicts the surface states with the Dirac cones on the boundary of the BZ. The similar effect was considered for the Bi$_{1-x}$Sb$_{x}$ alloy in Ref.~\cite{Teo-2008}. In Sec.~\ref{sec:continuous}, we discuss whether a continuous approximation for the TI Hamiltonian gives a good description of the surface states. We conclude that the tight-binding models are better suitable for the description of the surface states. Then, in Sec.~\ref{sec:generalizedShockley}, we generalize the Shockley model  by including additional tight-binding amplitudes. For all these models, we find that the edge state is always localized on one sublattice, which is rarely mentioned in the TI literature.

%%%%%%%%%%%%%%%%%%%%%%%%%%%%%%%%%%%%%%%%%%%%%%%%%%%%%%%%%%%%%%%%%%%%%%%%%%%%%
\section{1D Shockley model}  \label{sec:Shockley}
%%%%%%%%%%%%%%%%%%%%%%%%%%%%%%%%%%%%%%%%%%%%%%%%%%%%%%%%%%%%%%%%%%%%%%%%%%%%%

%%%%%%%%%%%%%%%%%%%%%%%%%%%%%%%%%%%%%%%%%%%%%%%%%%%%%%%%%%%%%%%%%%%%%%%%%%%%%
\subsection{The original Shockley model} \label{sec:classicalSchockley}
%%%%%%%%%%%%%%%%%%%%%%%%%%%%%%%%%%%%%%%%%%%%%%%%%%%%%%%%%%%%%%%%%%%%%%%%%%%%%

In this section, we briefly review the Shockley  model \cite{Shockley-1939,Davison-1996} and its properties. Let us consider  a 1D linear chain of atoms shown in Fig~\ref{fig:Shockley}(a).  The unit cell contains two atoms labeled as A and B, which are connected via the alternating nearest-neighbor complex tight-binding amplitudes $t_1$ and $t_2$. So, the Hamiltonian of the model is
\begin{eqnarray}
 &  H = \sum\limits_z \Psi^\dag(z)\left[U\Psi(z)+V\Psi(z-1)+V^\dag\Psi(z+1)\right], \label{Ham} \\
 &  U = \left(%
\begin{array}{cc}
  0 &  t_1^\ast \\
  t_1 & 0 \\
\end{array}%
\right),\,\,\,\,
V = \left(%
\begin{array}{cc}
  0 &  t_2^\ast \\
  0 & 0 \\
\end{array}%
\right). \label{uv0}
\end{eqnarray}
Here, $z$ is the integer coordinate of the unit cell, $t_1$ and $t_2$ are the intra-cell and the inter-cell tunneling amplitudes, and $\Psi(z)$ is the spinor
\begin{equation}
 \Psi(z) = \left(%
\begin{array}{c}
  \psi_a(z) \\
  \psi_b(z) \\
\end{array}%
\right), \label{spinor}
\end{equation}
where $\psi_a(z)$ and $\psi_b(z)$ are the wave functions on the sites $A$ and $B$. In the Fourier representation $\Psi(z) = \int_0^{2\pi}\frac{dk}{2\pi} \,e^{ikz}\, \Psi(k)$,  the Hamiltonian is
\begin{equation}
   H=\int_0^{2\pi} \frac{dk}{2\pi}\,\Psi^\dag(k)\,H(k)\,\Psi(k),
\end{equation}
where
\begin{equation}
  H(k)=U+Ve^{-ik}+V^\dag e^{ik}=\left(%
\begin{array}{cc}
  0 & t^\ast(k) \\
  t(k) & 0 \\
\end{array}%
\right)\label{BlochHam}\\
\end{equation}
is a $2\times 2$ matrix acting in the AB sublattice space, and
\begin{equation}
  t(k) = t_1+t_2e^{ik}=t_1+t_2q,\qquad q=e^{ik}. \label{tk}
\end{equation}
Then, the Schr\"odinger equation
\begin{eqnarray}
 &&  \left(%
\begin{array}{cc}
  0 & t^\ast(k) \\
  t(k) & 0 \\
\end{array}%
\right)  \left(%
\begin{array}{c}
  \psi_a \\
  \psi_b \\
\end{array}%
\right)=E \left(%
\begin{array}{c}
  \psi_a \\
  \psi_b \\
\end{array}%
\right) \label{Schrodinger}
\end{eqnarray}
gives two particle-hole symmetric energy bands with the eigenvalues $E(k)$ and eigenfunctions $\Psi(k)$
\begin{eqnarray}
&& E(k)=\pm|t(k)|, \label{spectr} \\
&&  \Psi(k) = \frac{1}{\sqrt 2} \left(%
\begin{array}{c}
  e^{i\arg[t(k)]} \\
  \pm 1 \\
\end{array}%
\right).\label{wvfunc1}
\end{eqnarray}
The energy spectrum has a gap if $|t_1|\neq|t_2|$, as illustrated in Fig.~\ref{fig:Shockley}(b) for real $t_1$ and $t_2$. Notice that the bulk wave function~(\ref{wvfunc1}) has equal probabilities on  both sublattices. In contrast, as we shall see below, the wave function of an edge state is localized only on one sublattice.

\begin{figure}
 \includegraphics[width=\linewidth]{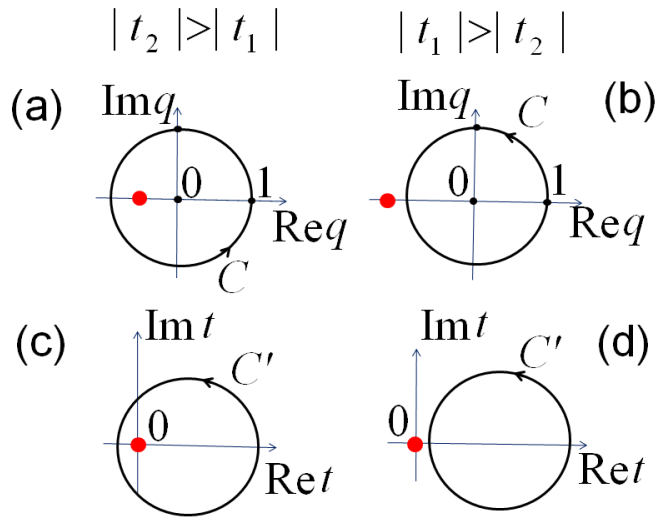}
 \caption{(Color online) Topological formulation of the Shockley criterion~(\ref{crit}).  Panels (a) and  (b) compare the two cases, where the root $q_0$ (the red dot) lies inside or outside the unit circle $C=\{q=e^{ik},\,k\in(0,2\pi)\}$. An edge state exists for $|q_0|<1$, panel (a), and does not exist for $|q_0|>1$, panel (b). An alternative formulation in terms of  the winding number~(\ref{winding}) is illustrated in panels (c) and (d). The edge state exists if the winding number is non-zero, panel (c), and does not exist if the winding number is zero, panel (d).} \label{fig:windingContours}
\end{figure}

A boundary to the 1D lattice can be introduced by cutting either $t_1$ or $t_2$ link. Let us consider a half-infinite system for $z\ge 1$, $z=1,\,2,\,3,\,\ldots$, corresponding to the cut of the $t_2$ link. In this case, the atom A is exposed on the edge, as shown in Fig.~\ref{fig:Shockley}(a). Mathematically, the boundary condition is introduced by requiring that the wave function vanishes at the fictitious  site $z=0$ and at infinity
\begin{eqnarray}
&&  \psi_a(0) = 0, \quad  \psi_b(0) = 0, \label{zbc} \\
&&  \psi_a(+\infty) = 0, \quad  \psi_b(+\infty) = 0. \label{ibc}
\end{eqnarray}
It is shown in Appendix~\ref{sec:appendixA} that the edge state can exist only for $E=0$. So, we substitute $E=0$ into Eq.~(\ref{Schrodinger}) and find that the wave functions on the $A$ and $B$ sublattices decouple
\begin{eqnarray}
  && t(k)\,\psi_a=(t_1+t_2e^{ik})\,\psi_a=0, \label{A}\\
  && t^\ast(k)\,\psi_b=(t_1^\ast+t_2^\ast e^{-ik})\,\psi_b=0,\label{B}
\end{eqnarray}
where $k$ is now a complex wave-number, so $t^\ast(k)$ is not a complex conjugate of $t(k)$. In the real space, Eq.~(\ref{B}) can be written as a recursion relation
\begin{equation}
\psi_b(z)t_1^\ast+\psi_b(z-1)t_2^\ast=0
\end{equation}
for $z\ge 1$. Using this recursion relation and the boundary condition $\psi_b(0)=0$, we find that $\psi_b(z)$ vanishes for $z\ge 1$. In contrast, the real-space representation of Eq.~(\ref{A})
\begin{equation}
\psi_a(z)t_1+\psi_a(z+1)t_2=0
\end{equation}
for $z\ge 1$ does not involve $\psi_a(0)$ from Eq.~(\ref{zbc}). So, the solution on the A sublattice is
\begin{equation}
\psi_a(z)=q_0^{z-1},\label{ES}
\end{equation}
where, $q_0$ is obtained by solving the equation $t(k_0)=0$, following from Eq.~(\ref{A}) for a complex wave-number $k_0$
\begin{eqnarray}
  q_0 = e^{ik_0}= -\frac{t_1}{t_2}. \label{qq}
\end{eqnarray}
Depending on whether  $|q_0|<1$ or $|q_0|>1$, the solution in Eq.~(\ref{ES}) either satisfies the condition~(\ref{ibc}) at infinity or not. If $|t_2|>|t_1|$, then $|q_0|<1$, as shown in Fig.~\ref{fig:windingContours}(a), and the wave function~(\ref{ES}) exponentially decays at $z\rightarrow +\infty$, as shown in Fig.~\ref{fig:Shockley}(c), so the edge state exists. In contrast, if $|t_1|>|t_2|$, then $|q_0|>1$, as shown in Fig.~\ref{fig:windingContours}(b), and the wave function~(\ref{ES}) exponentially grows at $z\rightarrow +\infty$, so an edge state does not exist. To summarize,  by solving Eqs.~(\ref{A})~and~(\ref{B}) with the appropriate boundary conditions (\ref{zbc}) and (\ref{ibc}), we obtain the zero-energy edge state
\begin{equation}
  \Psi_0(z) = \left(%
\begin{array}{c}
  1 \\
  0 \\
\end{array}%
\right)q_0^{z-1},\qquad E_0 = 0\, ,
\label{psi}
\end{equation}
which exists only if
\begin{equation}
|q_0|=\frac{|t_1|}{|t_2|}<1.  \label{crit}
\end{equation}
Equation~(\ref{crit}) constitutes the Shockley Criterion: {\it In the 1D tight-binding model with alternating tunneling amplitudes given by Hamiltonian~(\ref{Ham}), the edge state exists if the bond of the greater magnitude is broken at the boundary.}

%%%%%%%%%%%%%%%%%%%%%%%%%%%%%%%%%%%%%%%%%%%%%%%%%%%%%%%%%%%%%%%%%%%%%%%%%%%%%
%%%%%%%%%%%%%%%%%%%%%%%%%%%%%%%%%%%%%%%%%%%%%%%%%%%%%%%%%%%%%%%%%%%%%%%%%%%%%

Let us now consider an alternative formulation of the Shockley criterion~(\ref{crit}) in terms of the winding number
\begin{equation}
 W = \frac{1}{2\pi i}\int_0^{2\pi}dk\,\frac{d}{dk}\,{\rm ln}\,t(k).
 \label{winding}
\end{equation}
The winding number $W$ represents the phase change of the complex function $t(k)$ when the real variable $k$ changes from $0$ to $2\pi$. The  function $t(k)$ also defines a closed contour
\begin{equation}
 C'=\{t(k)=t_1+t_2e^{ik},\,k\in(0,2\pi)\} \label{cprime}
\end{equation}
in the 2D plane of $($Re$\,t,$Im$\,t)$, as shown in Fig.~\ref{fig:windingContours}, panels (c) and (d). If $|q_0|<1$, or equivalently $|t_2|>|t_1|$, the contour $C'$ winds around the origin (red dot), as shown in panel (c). If $|q_0|>1$, or equivalently $|t_1|>|t_2|$, the contour $C'$ does not wind around the origin, as shown in panel (d). So, the Shockley criterion~(\ref{crit}) can be formulated in terms of the winding number
\begin{equation}
 W = \left\{\begin{array}{ll}
   1,& {\rm \,\,edge\,\,state\,\, exists},  \\
   0,& {\rm \,\,edge\,\,state\,\,does\,\,not\,\,exist}.
 \end{array}\right. \label{critWinding}
\end{equation}
This formulation was discussed in a number of papers \cite{Ryu-2002,Mong-2011,Gurarie-2011,Deplace-2011}. While the winding number~(\ref{winding}) is calculated using the off-diagonal element $t(k)$ of the Hamiltonian~(\ref{BlochHam}), it can be equivalently expressed through the eigenfunctions $\Psi(k)$ defined in Eq.~(\ref{wvfunc1})
\begin{equation} 
  W_Z = \frac{1}{\pi i}\int_0^{2\pi}dk\,\Psi^\dag(k) \partial_k \Psi(k). \label{Zak}
\end{equation}
This expression is called the Zak phase \cite{Deplace-2011} (up to $\pi$ in the denominator and related to the Berry phase) and is an alternative representation of the winding number (\ref{winding}).

%%%%%%%%%%%%%%%%%%%%%%%%%%%%%%%%%%%%%%%%%%%%%%%%%%%%%%%%%%%%%%%%%%%%%%%%%%%%%
\subsection{On-site energies in the 1D Shockley model} \label{sec:shockleyOnSiteEnergy}
%%%%%%%%%%%%%%%%%%%%%%%%%%%%%%%%%%%%%%%%%%%%%%%%%%%%%%%%%%%%%%%%%%%%%%%%%%%%%
Let us further generalize the model and include on-site energies $\varepsilon_a$ and $\varepsilon_b$ in Hamiltonian~(\ref{BlochHam})
\begin{equation}
  H(k) = \left(%
\begin{array}{cc}
  \varepsilon_a & t^\ast(k) \\
  t(k) & \varepsilon_b \\
\end{array}%
\right),
\end{equation}
As shown in Eq.~(\ref{psi}) for $\varepsilon_a=\varepsilon_b=0$, the edge state solution is localized on the A sublattice. Therefore, adding the on-site energy $\varepsilon_a$ simply shifts the energy of the edge state without changing its wave function irrespective of $\varepsilon_b$. So, if criterion~(\ref{crit}) is satisfied, the edge state is localized on the $A$ sublattice and has the energy
\begin{equation}
E_0 = \varepsilon_a.
\end{equation}
It is also convenient to transform the Hamiltonian to the symmetrized form
 H\begin{equation}
   H(k) =\frac{\varepsilon_a+\varepsilon_b}{2} + \left(%
\begin{array}{cc}
  h & t^\ast(k) \\
  t(k) & -h \\
\end{array}%
\right),\,\,\, h=\frac{\varepsilon_a-\varepsilon_b}{2}.
\end{equation}
The offset $(\varepsilon_a+\varepsilon_b)/2$ just uniformly shifts all energies and will be omitted in the rest of the paper, so the Hamiltonian becomes
\begin{equation}
   H(k) = \left(%
\begin{array}{cc}
  h & t^\ast(k) \\
  t(k) & -h \\
\end{array}%
\right). \label{BlochHam4}
\end{equation}
The bulk spectrum of the Hamiltonian~(\ref{BlochHam4}) is generally gapped
\begin{equation}
  E(k)=\pm\sqrt{h^2+|t(k)|^2}.
\end{equation}

By denoting the Pauli matrices acting in the AB sublattice space as $\bm \tau = (\tau_x,\tau_y,\tau_z)$, Hamiltonian~(\ref{BlochHam4}) can be written as
\begin{equation}
  H (k) = \bm\tau\cdot\bm d(k), \qquad \bm d(k) = [{\rm Re}t(k),{\rm Im}t(k),h].
  \label{TauRepresent}
\end{equation}
When $k$ changes from $0$ to $2\pi$, the vector $\bm d(k)$ traces a closed contour $\Gamma$ in the corresponding 3D space. The criterion~(\ref{critWinding}) is equivalent to the following statement: The edge state exists if the projection of the contour $\Gamma$ onto the $xy$ plane encloses the origin \cite{Mong-2011}. Note that the Zak phase~(\ref{Zak}) is equal to $W_Z=\Omega/2\pi$, where $\Omega$ is the solid angle of the contour $\Gamma$ viewed from the origin. 
For $h=0$, the contour $\Gamma$ lies in the $xy$ plane, so $\Omega=2\pi$ and $W_Z=1$.  However, for $h\neq 0$, the contour $\Gamma$ lies off the $xy$ plane, and $\Omega$ is a fraction of $2\pi$. So, in general, the Zak phase $W_Z$ is fractional and does not give a number of the edge states, whereas the criterion~(\ref{critWinding}) remains applicable.

%%%%%%%%%%%%%%%%%%%%%%%%%%%%%%%%%%%%%%%%%%%%%%%%%%%%%%%%%%%%%%%%%%%%%%%%%%%%%
\section{3D Shockley-like model}  \label{sec:3DShockley}
%%%%%%%%%%%%%%%%%%%%%%%%%%%%%%%%%%%%%%%%%%%%%%%%%%%%%%%%%%%%%%%%%%%%%%%%%%%%%
%%%%%%%%%%%%%%%%%%%%%%%%%%%%%%%%%%%%%%%%%%%%%%%%%%%%%%%%%%%%%%%%%%%%%%%%%%%%%
\subsection{Generalization to the 3D case} \label{sec:1Dto3D}
%%%%%%%%%%%%%%%%%%%%%%%%%%%%%%%%%%%%%%%%%%%%%%%%%%%%%%%%%%%%%%%%%%%%%%%%%%%%%

%%%%%%%%%%%%%%%%%%%%%%%%%%%%%%%%%%%%%%%%%%%%%%%%%%%%%%%%%%%%%%%%%%%%%%%%%%%%%
\begin{figure}
 \includegraphics[width=\linewidth]{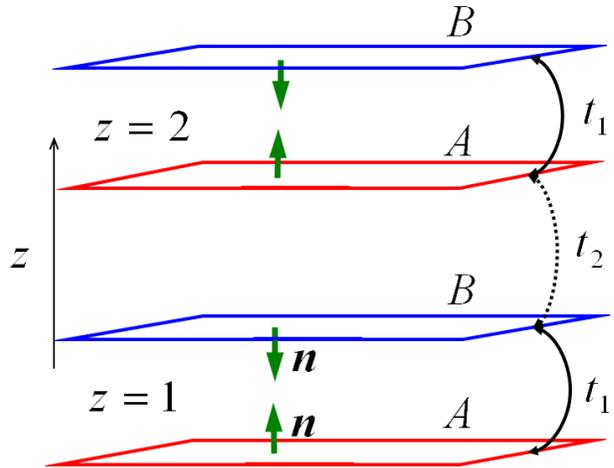}
\caption{(Color online) 3D generalization of the Shockley model described by Hamiltonian (\ref{BlochHam5}) with $h(\bm p)$ defined by Eq.~(\ref{dirac}). The arrows show the staggered direction of the Rashba vector $\bm n$.} \label{fig:layers}
\end{figure}
%%%%%%%%%%%%%%%%%%%%%%%%%%%%%%%%%%%%%%%%%%%%%%%%%%%%%%%%%%%%%%%%%%%%%%%%%%%%%

Let us generalize Hamiltonian~(\ref{BlochHam4}) to the 3D case.  Instead of alternating atomic sites, let us consider a sequence of alternating layers $A$ and $B$ perpendicular to the $z$ direction, as shown on Fig.~\ref{fig:layers}. Now, all parameters of Hamiltonian~(\ref{BlochHam4}) acquire dependence on the in-plane momentum $\bm p = (p_x,p_y)$
\begin{equation}
  H = \left(%
\begin{array}{cc}
  h(\bm p) & t^\ast(k,\bm p) \\
  t(k,\bm p) & -h(\bm p) \\
\end{array}%
\right). \label{BlochHam5}
\end{equation}
The off-diagonal matrix element
\begin{equation}
  t(k,\bm p)=t_1(\bm p)+t_2(\bm p)e^{ik} \label{complexFunction}
\end{equation}
describes the $\bm p$-dependent inter-layer tunneling amplitudes, while $h(\bm p)$ represents the intra-layer Hamiltonian.  Throughout this paper, we denote the in-plane momentum as $\bm p=(p_x,p_y)$ and the out-of-plane momentum in the $z$ direction as $k$ \cite{kpNote}.

For a fixed value of the in-plane momentum $\bm p$, Hamiltonian~(\ref{BlochHam5}) reduces to the 1D model~(\ref{BlochHam4}), for which the edge state was studied in Sec.~\ref{sec:Shockley}. The surface states exist for those in-plane momenta $\bm p$ where criterion~(\ref{crit}) is satisfied.  The surface states are localized on the A sublattice, and the energy spectrum $E_0(\bm p)$ of the surface states is determined by the in-plane Hamiltonian $h(\bm p)$
\begin{equation}
  E_0(\bm p) = h(\bm p). \label{eqspectr}
\end{equation}

In our construction of the generalized Shockley model, we put a restriction that the diagonal element $h(\bm p)$ does not depend on $k$.  Physically, it means that tunneling amplitudes connect only different sublattices A and B, but not A to A or B to B.  Thus, Hamiltonian~(\ref{BlochHam5}) is not the most general 3D Hamiltonian, however it applies to many models in the literature.

%%%%%%%%%%%%%%%%%%%%%%%%%%%%%%%%%%%%%%%%%%%%%%%%%%%%%%%%%%%%%%%%%%%%%%%%%%%%%
\subsection{Spin-orbit interaction} \label{sec:SpinOrbit}
%%%%%%%%%%%%%%%%%%%%%%%%%%%%%%%%%%%%%%%%%%%%%%%%%%%%%%%%%%%%%%%%%%%%%%%%%%%%%
\begin{figure}
 \includegraphics[width=\linewidth]{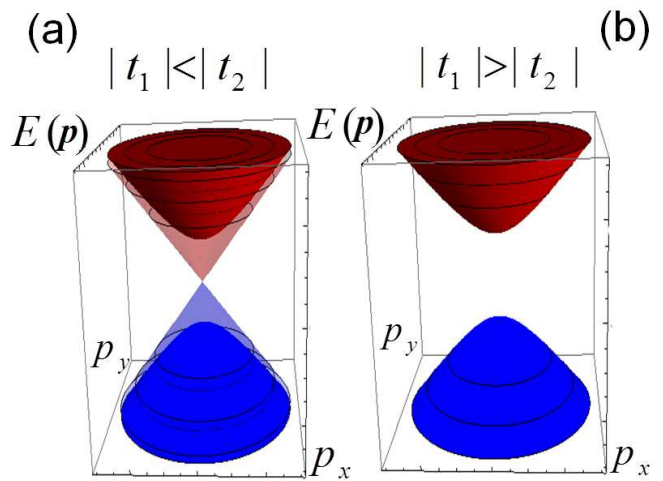}
\caption{(Color online)  Energy spectrum of the 3D Shockley model described by Hamiltonian~(\ref{BlochHam5}) in the vicinity of $\bm p=0$. The spectrum of the bulk states, Eq.~(\ref{spectrum1}), is shown by the solid parabolas in both panels. According to the Shockley criterion,  surface states exist if $|t_1|<|t_2|$, panel (a), and do not exist otherwise, panel (b). The surface states have the linear dispersion, Eq.~(\ref{diracSpectrum}), shown by the transparent Dirac cone in panel (b). } \label{fig:spectrum3D}
\end{figure}
%%%%%%%%%%%%%%%%%%%%%%%%%%%%%%%%%%%%%%%%%%%%%%%%%%%%%%%%%%%%%%%%%%%%%%%%%%%%%

So far, we have not considered spin of the electron.  After including the spin variable in Hamiltonian~(\ref{BlochHam5}), the terms $h(\bm p)$ and $t(k,\bm p)$ become $2\times2$ matrices acting in the spin-$1/2$ space, and the full Hamiltonian becomes a 4$\times$4 matrix. We assume that $t(k,\bm p)$ is proportional to the unit $2\times 2$  matrix, but $h(\bm p)$ may include the Pauli matrices $\bm \sigma$ acting on the spin variable.  In vicinity of the time-reversal-invariant momentum point $\bm p = 0$, the Hamiltonian $h(\bm p)$ must be bilinear in $\bm p$ and the spin-Pauli matrices $\bm \sigma$. For example, $h(\bm p)$ can have the Rashba spin-orbit coupling form
\begin{equation}
  h(\bm p)= v (\sigma_x p_y-\sigma_y p_x)=v(\bm\sigma\times\bm p)\cdot\hat z, \label{dirac}
\end{equation}
where $v$ has the dimension of velocity. Notice that the diagonal term $\pm h(\bm p)$ in Hamiltonian~(\ref{BlochHam5}) has opposite signs on the A and B sublattices. This corresponds to staggered direction of the Rashba vector $\bm n=\pm \hat{z}$ on different layers for the spin-orbit coupling $v\bm n(\bm\sigma\times\bm p)$ as shown in Fig.~\ref{fig:layers}. In the vicinity of $\bm p=0$, let us also approximate $t_1(\bm p) \approx t_1(0)$ and $t_2(\bm p)\approx t_2(0)$ and assume that $|t_1(0)|\neq |t_2(0)|$. Then, the surface states exist only if $|t_1(0)|<|t_2(0)|$, and the spectrum of the surface states has linear dependence on $|\bm p|$
\begin{equation}
  E_0(\bm p)=\pm v|\bm p|, \label{diracSpectrum}
\end{equation}
which is illustrated by the Dirac cone in panel~(a) of Fig.~\ref{fig:spectrum3D}. The wave functions of the surface states have in-plane spin-polarization perpendicular to the momentum $\bm p$. On the other hand, the bulk spectrum is parabolic in the  vicinity of $\bm p=0$
\begin{equation}
  E^2(k,\bm p) = |t(k,0)|^2+v^2\bm p^2, \label{spectrum1}
\end{equation}
as shown in both panels of Fig.~\ref{fig:spectrum3D} by  solid colors. Note that, because of the assumption $|t_1(0)|\neq |t_2(0)|$, the off-diagonal element $t(k,\bm p)$ is non-zero in the vicinity of $\bm p = 0$ and so the bulk spectrum~(\ref{spectrum1}) is gapped. On the other hand, if $|t_1(0)|=|t_2(0)|$, the bulk spectrum is gapless, and the Hamiltonian undergoes the topological phase transition, as will be shown in Sec.~\ref{sec:Diamond}.
%%%%%%%%%%%%%%%%%%%%%%%%%%%%%%%%%%%%%%%%%%%%%%%%%%%%%%%%%%%%%%%%%%%%%%%%%%%%%
\subsection{Vortex lines in 3D momentum space} \label{sec:VortexLines}
%%%%%%%%%%%%%%%%%%%%%%%%%%%%%%%%%%%%%%%%%%%%%%%%%%%%%%%%%%%%%%%%%%%%%%%%%%%%%

%%%%%%%%%%%%%%%%%%%%%%%%%%%%%%%%%%%%%%%%%%%%%%%%%%%%%%%%%%%%%%%%%%%%%%%%%%%%%
\begin{figure}
 \includegraphics[width=\linewidth]{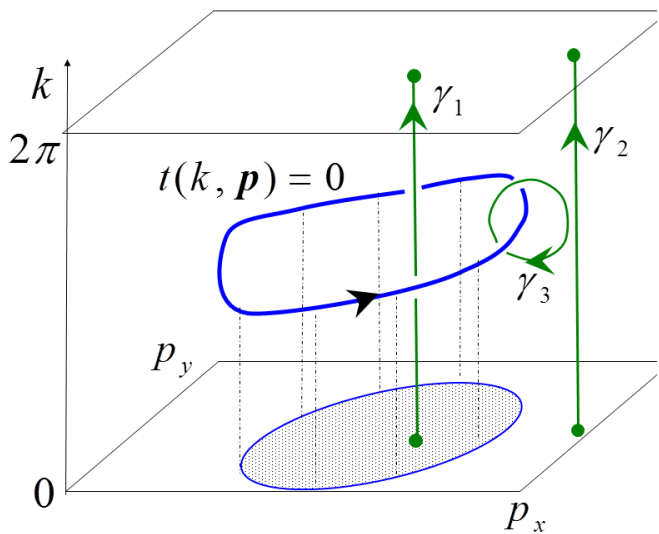}
 \caption{(Color online) The thick blue curve is a vortex line in the 3D momentum space defined by Eq.~(\ref{tkp0}). Its projection onto the 2D momentum space $\bm p$ defines the boundary of the shaded area, where the surface states exist.} \label{fig:nodal}
\end{figure}
%%%%%%%%%%%%%%%%%%%%%%%%%%%%%%%%%%%%%%%%%%%%%%%%%%%%%%%%%%%%%%%%%%%%%%%%%%%%%
In principle, the tunneling amplitudes $t_1(\bm p)$ and $t_2(\bm p)$ may depend on the in-plane momentum $\bm p$. So, the surface state existence criterion can only be satisfied in a certain domain of the 2D momentum space $\bm p$. In this section, we discuss how to identify this domain for the Hamiltonian~(\ref{BlochHam5}).

Let us consider the equation
\begin{equation}
  t(k,\bm p)=0 \label{tkp0}
\end{equation}
for the complex function $t(k,\bm p)$ in Eq.~(\ref{complexFunction}). It is equivalent to two equations Re~$t(k,\bm p)=0$ and Im~$t(k,\bm p)=0$, which define a line in the 3D momentum space $(k,\bm p)$. In general, the complex-valued function $t(k,\bm p)$ has a phase circulation around the line where it vanishes, i.e. Eq.~(\ref{tkp0}) defines a vortex line in the 3D momentum space~\cite{Beri-2010,Schnyder-2010,Heikkila-2011}. As an example, such a vortex line and its projection on the 2D momentum space $\bm p$  are shown in Fig.~\ref{fig:nodal}. Phase winding of the function $t(k,\bm p)$ along an arbitrary contour $\gamma$ can be calculated as
\begin{equation}
  W(\gamma) = \frac{1}{2\pi i}\oint_\gamma d\bm l\,\frac{d}{d\bm l}\,{\rm ln}\,t(k,\bm p),
 \label{winding2}
\end{equation}
 where the notation $\bm l=(k,\bm p)$ is used for brevity. For instance, the phase winding along the contour $\gamma_3$ around the vortex line in Fig.~\ref{fig:nodal} is non-zero
\begin{equation}
  W(\gamma_3)=1. \label{gamma3}
\end{equation}

Because the BZ is periodic in $k$, we can also define a closed contour by varying $0<k<2\pi$ for a fixed value of the in-plane momentum $\bm p$.  Such contours $\gamma_1$ and $\gamma_2$ are shown in Fig.~\ref{fig:nodal}, and the phase windings~(\ref{winding2}) are well defined for these contours. The contours $\gamma_1$ and $-\gamma_2$ can be merged into the contour $\gamma_3$. So, the following equation holds
\begin{equation}
  W(\gamma_3) = W(\gamma_1)-W(\gamma_2).
\end{equation}
Given Eq.~(\ref{gamma3}) and the condition~(\ref{critWinding}) that $W(\gamma_{1,2})\ge 0$, we find that the winding numbers  are $W(\gamma_1)=1$ and $W(\gamma_2)=0$.  Since a non-zero winding number is required for existence of the surface states according to Eq.~(\ref{critWinding}), we conclude that the surface states exist for the 2D momenta $\bm p$ in the shaded area of Fig.~\ref{fig:nodal} and do not exist outside. Thus, we have shown that the projection of the vortex line~(\ref{tkp0}) onto the 2D momentum space  $\bm p$ defines the domain where the surface states exist.

While the main focus of this work is the 3D systems, let us comment on the 2D case $\bm l=(k,p_x)$, where $p_x$ and $k$ are the momenta parallel and perpendicular to the edge of the 2D system. The 2D case can also be viewed as a slice of 3D momentum space shown in Fig.~\ref{fig:nodal} at a fixed momentum $p_y$. Then, the solution of the equation $t(k,p_x)=0$ generally defines a set of vortex points in the 2D momentum space $\bm l$. Similarly to the 3D case, a projection of the vortex points onto the $p_x$ momentum space identifies a domain in $p_x$ for which the edge states exist. This method was used in Ref.~\cite{Deplace-2011} to find the edge states in graphene ribbons.
%%%%%%%%%%%%%%%%%%%%%%%%%%%%%%%%%%%%%%%%%%%%%%%%%%%%%%%%%%%%%%%%%%%%%%%%%%%%
\section{Diamond Model} \label{sec:Diamond}
%%%%%%%%%%%%%%%%%%%%%%%%%%%%%%%%%%%%%%%%%%%%%%%%%%%%%%%%%%%%%%%%%%%%%%%%%%%%%
%%%%%%%%%%%%%%%%%%%%%%%%%%%%%%%%%%%%%%%%%%%%%%%%%%%%%%%%%%%%%%%%%%%%%%%%%%%%%
\subsection{Hamiltonian and surface states} \label{sec:DiamondHamiltonian}
%%%%%%%%%%%%%%%%%%%%%%%%%%%%%%%%%%%%%%%%%%%%%%%%%%%%%%%%%%%%%%%%%%%%%%%%%%%%%
%%%%%%%%%%%%%%%%%%%%%%%%%%%%%%%%%%%%%%%%%%%%%%%%%%%%%%%%%%%%%%%%%%%%%%%%%%%%%
\begin{figure}
 \includegraphics[width=\linewidth]{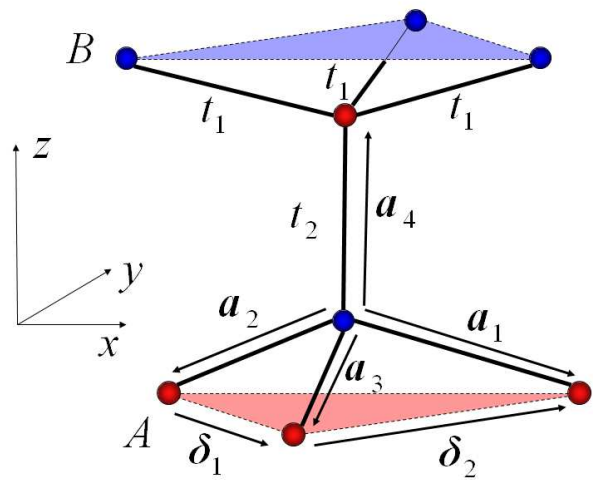}
\caption{(Color online)  Illustration of the diamond crystal structure and the tight-binding model described by Hamiltonian~(\ref{fullHam}). The lattice has two atoms in a unit cell shown by the red (A) and blue (B) spheres. } \label{fig:diamondStruct}
\end{figure}
%%%%%%%%%%%%%%%%%%%%%%%%%%%%%%%%%%%%%%%%%%%%%%%%%%%%%%%%%%%%%%%%%%%%%%%%%%%%%
%%%%%%%%%%%%%%%%%%%%%%%%%%%%%%%%%%%%%%%%%%%%%%%%%%%%%%%%%%%%%%%%%%%%%%%%%%%%%
\begin{figure}
 \includegraphics[width=\linewidth]{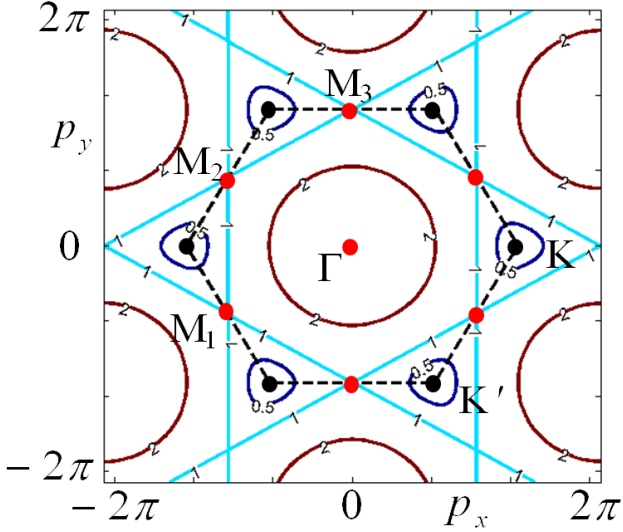}
\caption{(Color online) Lines of constant value for the graphene spectrum function $|t_1(\bm p)|/|t_1| = C$, for $C=0.5,\,1,\,2$, within the Brillouin zone (BZ), denoted by the dashed lines. The contour lines degenerate to points at the BZ corners (K and K$'$ points)  at $C=0$ and at the BZ center at $C=3$. The thick red dots denote the time-reversal-invariant momenta points~(\ref{trim}).} \label{fig:contours2D}
\end{figure}
%%%%%%%%%%%%%%%%%%%%%%%%%%%%%%%%%%%%%%%%%%%%%%%%%%%%%%%%%%%%%%%%%%%%%%%%%%%%%

In this section, we illustrate how the Shockley model can be applied to study the surface states for a particular TI model of Ref.~\cite{Kane-2007}. However similar approach can be applied to other models \cite{Kane-2005,Bernevig-2006,Hosur-2010}.

Let us consider a tight-binding model on the diamond lattice shown in Fig.~\ref{fig:diamondStruct}. The diamond lattice has two equivalent atom positions denoted by A (red) and B (blue). Atoms of each type form 2D triangular lattices, so that the A and B layers alternate along the $z$ direction similarly to Fig.~\ref{fig:layers}. The nearest A and B layers form a distorted graphene lattice. So, when viewed along the z direction, the structure looks like the ABC-stacked graphite lattice. We define the nearest-neighbor vectors $\bm a_{n}$, $n=1,2,3,4$, as shown in Fig.~\ref{fig:diamondStruct}, as well as the vectors
\begin{eqnarray}
&   \bm \delta_1 = \bm a_3-\bm a_2=( 1/2\,,\, -\sqrt{3}/2 ), \nonumber\\
&   \bm \delta_2 = \bm a_1-\bm a_3=(1/2\,,\, \sqrt{3}/2), \label{deltas}\\
&   \bm \delta_3 = \bm a_2-\bm a_1=(-1\,,\, 0), \nonumber
\end{eqnarray}
 which are the in-plane elementary translation vectors of the unit length $|\bm\delta_{n}|=1$.

The Hamiltonian of the model has the form of Eq.~(\ref{BlochHam5})
\begin{equation}
H = \left(%
\begin{array}{cc}
  h(\bm p) & t^\ast(k,\bm p) \\
  t(k,\bm p) & -h(\bm p) \\
\end{array}
\right), \label{fullHam}
\end{equation}
where the unit cell consists of the A and B atoms connected by the vector $\bm a_3$. The off-diagonal part
\begin{eqnarray}
&& t(k, \bm p) =  t_1(\bm p)+t_2 e^{ik}, \label{tkp} \\
&& t_1(\bm p) = t_1(1+e^{-i\bm p\bm \delta_1}+e^{i\bm p\bm \delta_2}), \label{fp}
\end{eqnarray}
describes the nearest-neighbor tunneling between the A and B sublattices with the amplitude $t_1$ along the vectors $\bm a_{n}$, $n=1,2,3$,  and the amplitude $t_2$ along the vector $\bm a_4$ \cite{kNote}. In Eqs.~(\ref{tkp}) and (\ref{fp}), we distinguish between the in-plane-momentum-dependent function $t_1(\bm p)$ and the tight-binding amplitude $t_1$.  Equation~(\ref{fp}) describes the well-known tight-binding spectrum of graphene \cite{CastroNeto-2009}
\begin{equation}
|t_1(\bm p)|/|t_1|=\sqrt{3+2\cos(\bm p\bm\delta_1)+2\cos(\bm p\bm\delta_2)+2\cos(\bm p\bm\delta_3)}
\end{equation}
The contour plots of $|t_1(\bm p)|/|t_1|=C$, for $C=0.5,\,1,\,2$, are shown in Fig.~\ref{fig:contours2D}. Note that $t_1(\bm p)$ has the linear Dirac-like dependence on the momentum $\bm p$ at the BZ corners, K and K$'$ points in Fig.~\ref{fig:contours2D}.

\begin{figure}
 \includegraphics[width=\linewidth]{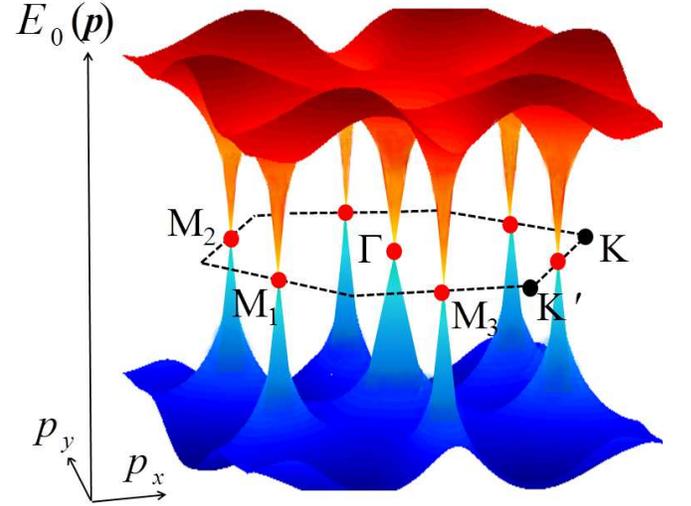}
\caption{(Color online)  The plot of the particle-hole symmetric spectrum $E_0(\bm p)$, Eq.~(\ref{spectr2}), induced by the spin-orbit Hamiltonian $h(\bm p)$~(\ref{hp}). In the vicinity of the time reversal invariant points, shown with the thick red dots, the Hamiltonian~(\ref{hp}) becomes linear in momentum. The dashed line denotes the boundary of the BZ.} \label{fig:spinStates}
\end{figure}

The diagonal term $h(\bm p)$ in Hamiltonian~(\ref{fullHam}) describes the spin-orbit interaction~\cite{Kane-2007}
\begin{equation}
  h(\bm p) = \frac{2\sqrt{2}}{3}\Lambda_{\rm SO}\sum_{i,j,l=1,2,3}
\epsilon_{ijl}\left(\bm \sigma \cdot
[\bm a_i\times\bm a_j]\right) \sin (\bm p \bm \delta_l),
\label{hp}
\end{equation}
where $\Lambda_{\rm SO}$ is the strength of the spin-orbit coupling, and $\epsilon_{ijl}$ is the antisymmetric tensor.  For simplicity, we do not include the inter-layer spin-orbit coupling involving the vector $\bm a_4$ in Hamiltonian~(\ref{hp}), unlike in Ref.~\cite{Kane-2007}. Hamiltonian~(\ref{hp}) has a gapless particle-hole symmetric spectrum $E_0(\bm p) = \pm \sqrt{h^2(\bm p)}$,
\begin{eqnarray}
  && E_0^2(\bm p)/\Lambda_{\rm SO}^2 =h^2(\bm p)/\Lambda_{\rm SO}^2= \sin^2 (\bm p \bm \delta_1)+  \label{spectr2} \\
  && = \sin^2 (\bm p \bm \delta_2)+  \sin^2 (\bm p \bm\delta_3) +\sin (\bm p \bm \delta_1)\sin (\bm p \bm \delta_2)+\nonumber\\
  && + \sin (\bm p \bm \delta_1)\sin (\bm p \bm  \delta_3)+\sin (\bm p \bm \delta_2)\sin (\bm p \bm  \delta_3),\nonumber
\end{eqnarray}
which is shown in Fig.~\ref{fig:spinStates}.
%%%%%%%%%%%%%%%%%%%%%%%%%%%%%%%%%%%%%%%%%%%%%%%%%%%%%%%%%%%%%%%%%%%%%%%%%%%%%
%%%%%%%%%%%%%%%%%%%%%%%%%%%%%%%%%%%%%%%%%%%%%%%%%%%%%%%%%%%%%%%%%%%%%%%%%%%%%
The energy $E_0(\bm p)$ vanishes at the four time-reversal-invariant momenta (TRIM)
\begin{equation}
 \bm p_\ast\in \{\Gamma,M_1, M_2, M_3\},
\label{trim}
\end{equation}
where $\Gamma$ is the BZ center, and $M_1,\, M_2,\, M_3$ are at the centers of the BZ boundary, as shown in Figs.~\ref{fig:contours2D}~and~\ref{fig:spinStates}. Hamiltonian~(\ref{hp}) is bilinear in the momentum and spin operators in the vicinity of the TRIM points
\begin{eqnarray}
  &&  \frac{h(\bm p+\bm p_\ast^{(\Gamma)})}{\Lambda_{\rm SO}}\approx\frac{\sqrt{3}}{2}(\sigma_x p_y-\sigma_y p_x),  \label{dirac1}\\
  &&  \frac{h(\bm p+\bm p_\ast^{(M_3)})}{\Lambda_{\rm SO}}\approx-\frac{\sqrt{3}}{2}\sigma_x p_y-\frac{1}{2\sqrt{3}}\sigma_y
 p_x-\frac{2\sqrt{2}}{\sqrt{3}}\sigma_z p_x. \nonumber\\ \label{dirac2}
\end{eqnarray}
Thus, the energy spectrum $E_0(\bm p)$ has the shape of the Dirac cones in the vicinity of TRIM points, as shown in Fig.~\ref{fig:spinStates}. It is important to distinguish the linear, Dirac-like, behavior of the off-diagonal term $t_1(\bm p)$ in the vicinity of the BZ corners (K and K$'$ points) and of the diagonal term $h(\bm p)$ in the vicinity of the TRIM points, which are different sets of points in the BZ.

The bulk spectrum of Hamiltonian~(\ref{fullHam})
\begin{eqnarray}
  && E^2(k,\bm p) = |t(k,\bm p)|^2+E_0^2(\bm p),\label{bulk}
\end{eqnarray}
contains contributions from both the diagonal $h(\bm p)$ and the off-diagonal $t(k,\bm p)$ terms. The bulk spectrum becomes gapless when both contributions vanish for some momenta $(k,\bm p)$
\begin{eqnarray}
& E_0(\bm p)=0, \label{trimEq} \\
&  t(k,\bm p)=0. \label{vortexEq2}
\end{eqnarray}
Given Eq.~(\ref{tkp}), Eq.~(\ref{vortexEq2}) is equivalent to
\begin{equation}
 |t_1(\bm p)|=|t_2|, \label{vortexEq}
\end{equation}
which defines a contour line in Fig.~\ref{fig:contours2D}. Conditions~(\ref{trimEq})~and~(\ref{vortexEq2}) can be satisfied simultaneously only for special values of the parameters $t_1$ and $t_2$. The bulk spectrum becomes gapless, for $|t_2|=|t_1|$, when the contour line~(\ref{vortexEq}) passes through the TRIM points $M_1,\, M_2,\, M_3$, and for $|t_2|=3|t_1|$, when it passes through the $\Gamma$ point.

 %%%%%%%%%%%%%%%%%%%%%%%%%%%%%%%%%%%%%%%%%%%%%%%%%%%%%%%%%%%%%%%%%%%%%%%%%%%%%
\begin{table}
  \begin{tabular*}{0.45\textwidth}{@{\extracolsep{\fill}} c  c  c  c  }
 \hline\hline
                              & $t_2$ broken                      &  $ t_1$  broken                               & TI            \\ \hline
  $0<|t_2|<|t_1|$             & $-$                               & $M_{1,2,3},\,\Gamma$                          & Weak          \\
  $|t_1|<|t_2|<3|t_1|$        & $M_{1,2,3}$                       & $\Gamma$                                      & Strong        \\
  $3|t_1|<|t_2|$              & $M_{1,2,3},\,\Gamma$              & $-$                                           & Weak          \\ \hline\hline
\end{tabular*}
\caption{The table shows the points in the BZ where the surface states exist depending on the parameters of the model and which bond is broken at the surface. According to Fig.~\ref{fig:spinStates}, the surface states have the Dirac cones at the corresponding points. Letters $M_1,\, M_2,\, M_3,\, \Gamma$ denote positions of the TRIM points~(\ref{trim}).} \label{Table1}
\end{table}
%%%%%%%%%%%%%%%%%%%%%%%%%%%%%%%%%%%%%%%%%%%%%%%%%%%%%%%%%%%%%%%%%%%%%%%%%%%%

Hamiltonian~(\ref{fullHam}) has the Shockley form, Eq.~(\ref{BlochHam5}). Therefore all the conclusions of Secs.~\ref{sec:Shockley} and~\ref{sec:3DShockley} apply here, including the criterion~(\ref{crit}) for existence of the surface states. We find that the surface states have the dispersion $E_0(\bm p)$ and exist for those in-plane momenta $\bm p$ where the following condition is satisfied
\begin{equation}
  |t_1(\bm p)|<|t_2|. \label{Shockley1}
\end{equation}
The boundary of this domain is given by Eq.~(\ref{vortexEq}). When we change the parameter $t_2$ while keeping $t_1$ fixed, the Hamiltonian undergoes a transition between the phases with odd and even numbers of surface Dirac cones, called the ``strong'' and ``weak'' TI phases in Ref.~\cite{Kane-2007}. For small $|t_2|\ll |t_1|$, the contour lines given by Eq.~(\ref{vortexEq}) wind around the BZ corners K and K$'$ (see Fig.~\ref{fig:contours2D} for $C=0.5$), and criterion~(\ref{Shockley1}) is satisfied in the small area inside. The surface states do not include the Dirac cones of $E_0(\bm p)$, shown in Fig.~\ref{fig:spinStates}, because the TRIM points (red dots) are in the area where Eq.~(\ref{Shockley1}) is not satisfied. Thus, Hamiltonian~(\ref{fullHam}) is in the ``weak'' TI phase in this case. For $|t_1|=|t_2|$, the contours~(\ref{vortexEq}) become straight lines passing through the TRIM points $M_1,\,M_2,\,M_3$, as shown in Fig.~\ref{fig:contours2D} for $C=1$. So, both Eqs.~(\ref{trimEq})~and~(\ref{vortexEq}) are satisfied at the TRIM points, and the bulk spectrum, Eq.~(\ref{bulk}), becomes gapless. This marks a transition to the ``strong'' TI phase. When $|t_1|<|t_2|<3|t_1|$, the contour forms a circle around the BZ center, see Fig.~\ref{fig:contours2D} for $C=2$. Criterion Eq.~(\ref{Shockley1}) is satisfied in the exterior of the circle, and so the surface states contain the Dirac cones at the TRIM points $M_1$, $M_2$ and $M_3$. When $t_2$ reaches the critical value $|t_2|=3|t_1|$, the contour~(\ref{vortexEq}) shrinks to the single point $\Gamma$. The bulk spectrum becomes gapless, and this marks a transition to the ``weak'' TI phase again. For $|t_2|>3|t_1|$, the Shockley criterion is satisfied everywhere in the BZ, so the surface states include the Dirac cones for all TRIM points~(\ref{trim}).

As discussed above, the Shockley criterion~(\ref{Shockley1}) is written for the case where the $t_2$ bond is broken at the surface. If, on the other hand, the crystal termination is such that the $t_1$ bond is broken at the surface, the existence criterion for the surface state becomes complementary to the criterion~(\ref{Shockley1})
\begin{equation}
   |t_1(\bm p)|>|t_2|.
\end{equation}
So, the surface states now exist for those momenta $\bm p$ where they did not exist in the case of the broken bond $t_2$ and have the Dirac cones at the complementary TRIM points. This is summarized in Table~\ref{Table1}, which shows the Dirac cones belonging to the surface states depending on whether $t_1$ or $t_2$ is broken at the surface. In the ``strong'' TI phase, there is an odd number of the Dirac cones in the surface states, so, at least, one surface Dirac cone always exists. In contrast, in the ``weak'' TI phase, there is an even number of the surface Dirac cones, so the surface states may disappear under certain conditions.

%%%%%%%%%%%%%%%%%%%%%%%%%%%%%%%%%%%%%%%%%%%%%%%%%%%%%%%%%%%%%%%%%%%%%%%%%%%%%
\subsection{3D vortex lines} \label{sec:Vortex}
%%%%%%%%%%%%%%%%%%%%%%%%%%%%%%%%%%%%%%%%%%%%%%%%%%%%%%%%%%%%%%%%%%%%%%%%%%%%%

%%%%%%%%%%%%%%%%%%%%%%%%%%%%%%%%%%%%%%%%%%%%%%%%%%%%%%%%%%%%%%%%%%%%%%%%%%%%%
\begin{figure}
\centering
\begin{tabular}{c}
 \includegraphics[width=0.7\linewidth]{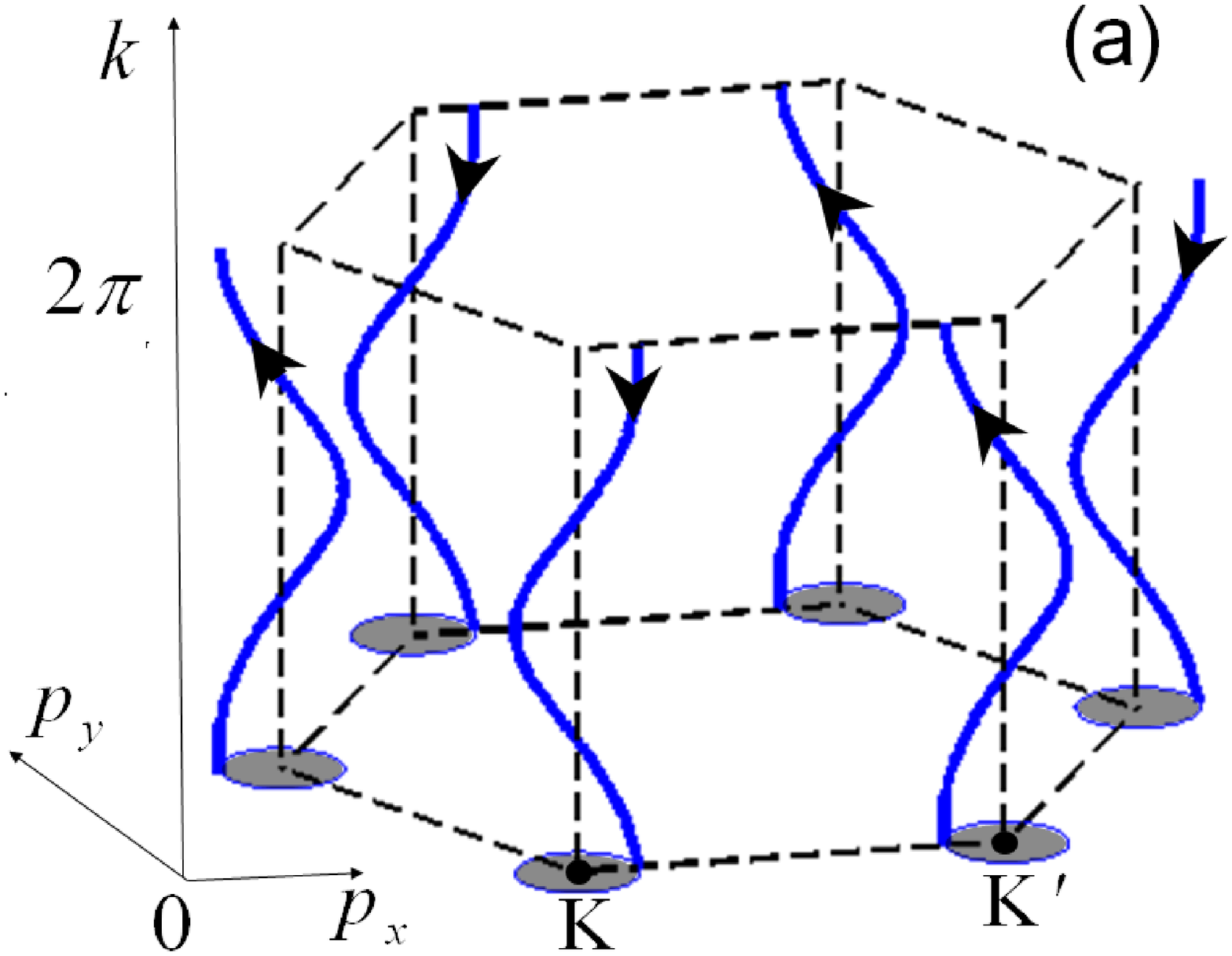} \\
\includegraphics[width=0.7\linewidth]{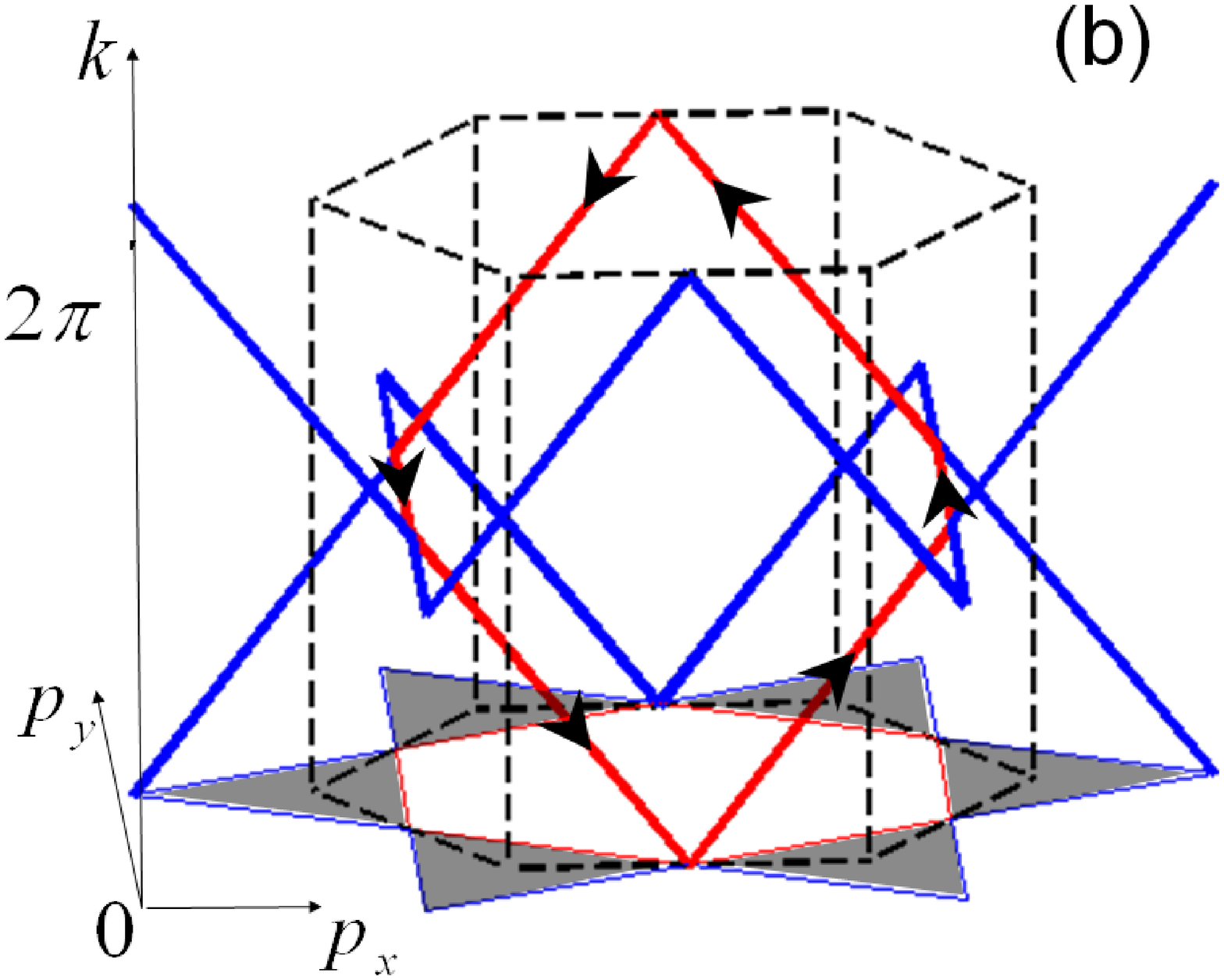} \\
\includegraphics[width=0.7\linewidth]{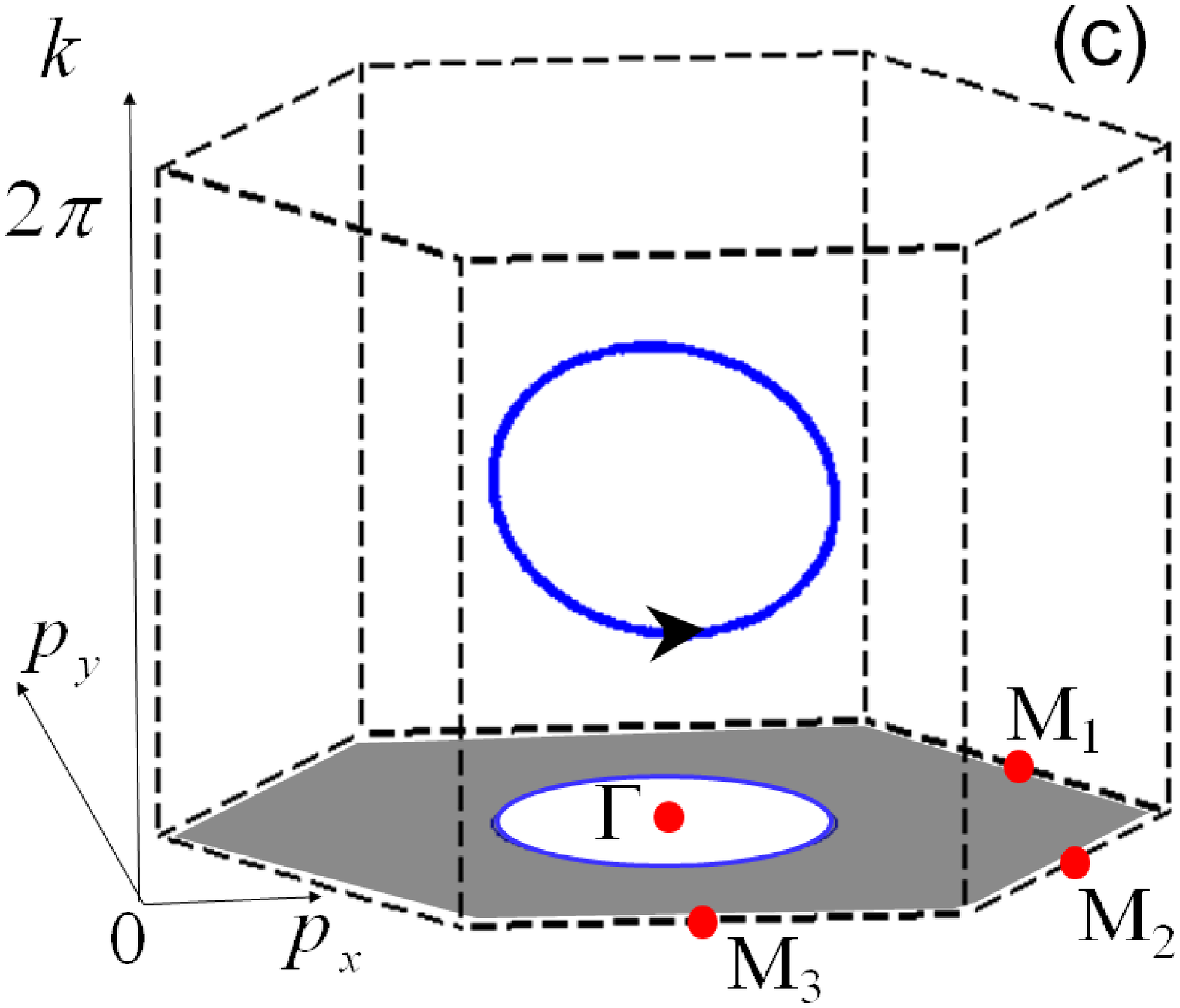}\\
\end{tabular}
\caption{(Color online) Vortex lines in the 3D momentum space defined by Eq.~(\ref{vortexEq1}), and shown for different values of the parameters: (a) $|t_1|>|t_2|$, (b) $|t_1|=|t_2|$, and (c) $|t_1|<|t_2|<3|t_1|$. The vortex lines are shown by the thick lines with the arrows representing vorticity. The thin lines show projections of the vortex lines, which encircle the shaded area in the 2D momentum space $\bm p$, where the Shockley criterion~(\ref{Shockley1}) is satisfied and the surface states exist. The dashed lines show the boundaries of the BZ. The part of the vortex lines residing in the first BZ is highlighted in red in panel (b). The three panels show the evolution of the vortex lines with the change of the parameters of the Hamiltonian. At $|t_1|=|t_2|$,  the vortex lines reconnect at the TRIM points and change their topology from spirals for $|t_1|>|t_2|$  to the loop for $|t_1|<|t_2|$. The change of the vortex lines topology is responsible for a transition from the ``weak'' to ``strong'' TI phase in the Hamiltonian~(\ref{fullHam}).}\label{fig:cont}
\end{figure}
%%%%%%%%%%%%%%%%%%%%%%%%%%%%%%%%%%%%%%%%%%%%%%%%%%%%%%%%%%%%%%%%%%%%%%%%%%%%%

In the previous section, we showed that the 2D contour defined by Eq.~(\ref{vortexEq}) represents the boundary separating the domain in the 2D momentum space where the Shockley criterion is satisfied.   On the other hand, the contour~(\ref{vortexEq}) is just the  projection of the 3D vortex line, defined by Eq.~(\ref{vortexEq2})
\begin{equation}
  t(k,\bm p)=t_1(\bm p)+t_2e^{ik}=0, \label{vortexEq1}
\end{equation}
onto the 2D momentum space, as discussed in Sec.~\ref{sec:3DShockley}. Let us discuss evolution of these 3D vortex lines with the change of the parameters $t_1$ and $t_2$. In the vicinity of the BZ corners $\bm p_0 = (\pm 4\pi/3,0)$, K and K$'$ points in Fig.~\ref{fig:contours2D}, where the function $t_1(\bm p)$ vanishes, Eq.~(\ref{fp}) can be linearized
\begin{equation}
  t_1(\bm p_0+\bm p)\approx-\frac{\sqrt{3}}{2}t_1(\pm p_x+ip_y). \label{t1a}
\end{equation}
So, for $|t_2|\ll |t_1|$, Eq.~(\ref{vortexEq1}) with $t_1(\bm p)$ defined in Eq.~(\ref{t1a}) describes  spirals in the 3D momentum space $(k,\bm p)$ \cite{Arovas-2008}
\begin{eqnarray}
 \left(%
\begin{array}{c}
  p_x \\
  p_y \\
\end{array}%
\right) = \frac{2}{\sqrt 3}\frac{t_2}{t_1}
\left(
\begin{array}{c}
   \pm \cos k  \\
    \sin k\\
 \end{array}\right),
\end{eqnarray}
as shown in Fig.~\ref{fig:cont}(a). Projections of these spirals onto the 2D momentum space $\bm p$ encircle the corners K and K$'$ of the 2D BZ. With the increase of $t_2$, the spirals grow until $t_2$ reaches the critical value $|t_2|=|t_1|$. At this point, the vortex lines reconnect as shown in Fig.~\ref{fig:cont}(b) and transform into three families of straight lines obtained by intersections of the planes
\begin{eqnarray}
  &&\{\bm{p}\bm\delta_1 = \pi+2\pi n\}\bigcap \{\bm{p}\bm\delta_3 = -k+2\pi
  m\},\\
  &&\{\bm{p}\bm\delta_2 = \pi+2\pi n\}\bigcap \{\bm{p}\bm\delta_3 =
k+2\pi
  m\},\\
  &&\{\bm{p}\bm\delta_3 = \pi+2\pi n\}\bigcap \{k =
  \pi+2\pi m\},
\end{eqnarray}
where $n$ and $m$ are independent integers. The part of these lines residing in the first BZ forms a loop highlighted in red for clarity in Fig.~\ref{fig:cont}(b). With the further increase of $t_2$, the vortex line detaches from the BZ boundary and becomes a closed loop, as shown in Fig.~\ref{fig:cont}(c).  In the vicinity of the $\Gamma$ point, the function $t_1(\bm p)$ given by Eq.~(\ref{fp}) can be expanded to the second order in $\bm p$, so the vortex line defined by Eq.~(\ref{vortexEq1}) is given by the intersection of a cylinder and a plane in the 3D momentum space $(k,\bm p)$
\begin{equation}
 \left\{  p_x^2 + p_y^2 = \frac{2}{3}\left(9 -\frac{t_2^2}{t_1^2}\right)\right\} \bigcap \left\{k = \pi+\sqrt 3\frac{t_1}{t_2}p_y\right\}.
\end{equation}
For the critical value $|t_2| = 3|t_1|$, the vortex line shrinks to the $\Gamma$ point and then disappears for $|t_2|>3|t_1|$.

So, we observe that the vortex lines  change their topology at the critical values of the model parameters $|t_2| = |t_1|$ and $|t_2|=3|t_1|$. These are the critical values where the transitions happen between the ``weak'' and ``strong'' TI phases. So, the configuration of the vortex lines (\ref{vortexEq1}) is directly related to the topological phase of the full Hamiltonian $H$, Eq.~(\ref{fullHam}).

Now, let us illustrate that the vortex lines are gauge-dependent, i.e., different choice of phases in the tight-binding model leads to different vortex lines. Let us choose the elementary cell consisting of the A and B atoms connected via the vector $\bm a_2$ shown in Fig.~(\ref{fig:diamondStruct}), rather than $\bm a_1$ chosen in Eqs.~(\ref{fullHam})-(\ref{fp}). Then, Eq.~(\ref{fp}) becomes $t_1(\bm p)=t_1\left(1+e^{-i\bm p\bm \delta_3}+e^{i\bm p \delta_1}\right)$, which is equivalent to the $2\pi/3$ rotation of $t_1(\bm p)$ in Eq.~(\ref{fp}) around the $k$ axis. Since $t_1(\bm p)$ defines the vortex lines via Eq.~(\ref{vortexEq1}), the vortex lines are $2\pi/3$ rotated compared to the lines shown in Fig.~\ref{fig:cont}. Notice, however, that the area where the surface states exist, shown by the shaded area in Fig.~\ref{fig:cont}, is $C_3$ symmetric and thus remains the same for a different gauge choice. 

We also point out that the Shockley Hamiltonian~(\ref{fullHam}) and the vortex lines are constructed for a particular crystal termination and cannot be directly used to study surface states for other surfaces. For a different crystal termination, we need to redefine the in-plane $\bm p'$ and the out-of-plane $k'$ momenta relative to the  ``new'' surface. Since the ``new'' momenta $(k',p'_x,p'_y)$ are related to the ``old'' momenta $(k,p_x,p_y)$ through some orthogonal transformation $O$: $(p_x,p_y,k)^{\rm T} = O\, (p'_x,p'_y,k')^{\rm T}$, the diagonal element of Hamiltonian~(\ref{fullHam}) is generally a function of both $k'$ and $\bm p'$: $h(\bm p) = h(k',\bm p')$. So, the Hamiltonian of the ``new'' surface does not have the Shockley form~(\ref{fullHam}), which requires that $h(\bm p')$ is independent of $k'$, and the Shockley criterion is not directly applicable (see a discussion in the end of Sec.~\ref{sec:1Dto3D}). 

%%%%%%%%%%%%%%%%%%%%%%%%%%%%%%%%%%%%%%%%%%%%%%%%%%%%%%%%%%%%%%%%%%%%%%%%%%%%%
\section{Shockley model description of $\mathbf{Bi_2Se_3}$} \label{sec:BiSeShockley}
%%%%%%%%%%%%%%%%%%%%%%%%%%%%%%%%%%%%%%%%%%%%%%%%%%%%%%%%%%%%%%%%%%%%%%%%%%%%%
%%%%%%%%%%%%%%%%%%%%%%%%%%%%%%%%%%%%%%%%%%%%%%%%%%%%%%%%%%%%%%%%%%%%%%%%%%%%%
\subsection{General analysis} \label{sec:BiSeGeneral}
%%%%%%%%%%%%%%%%%%%%%%%%%%%%%%%%%%%%%%%%%%%%%%%%%%%%%%%%%%%%%%%%%%%%%%%%%%%%%

%%%%%%%%%%%%%%%%%%%%%%%%%%%%%%%%%%%%%%%%%%%%%%%%%%%%%%%%%%%%%%%%%%%%%%%%%%%%%
\begin{figure}
\centering
\includegraphics[width=\linewidth]{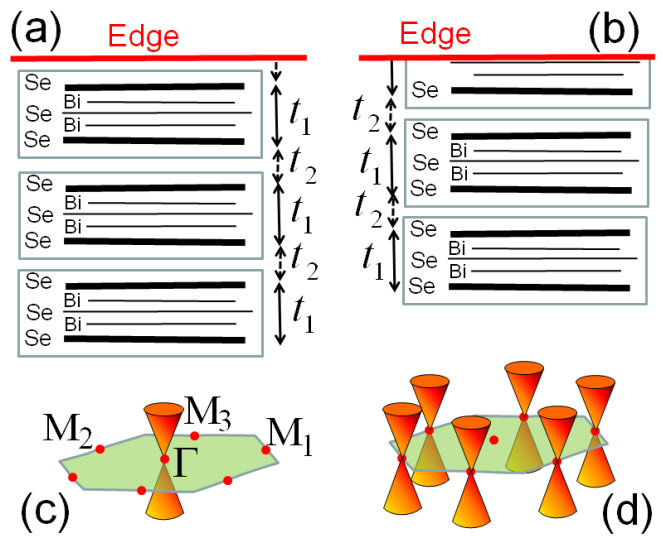}
\caption{(Color online) The crystal  of Bi$_2$Se$_3$ is formed by quintuple layers, schematically shown by the blue boxes. Each quintuplet  consists of the alternating layers Se-Bi-Se-Bi-Se. The tight-binding tunneling amplitudes $t_1$ and $t_2$ connect the orbitals of the outermost edges of the quintuplets. Then, depending on whether $t_2$ or $t_1$ is broken at the surface, as shown by the red line in panels (a) and (b), surface states occur in different regions of the 2D momentum space, as shown in Panels (c) and (d).}\label{fig:cuts}
\end{figure}
%%%%%%%%%%%%%%%%%%%%%%%%%%%%%%%%%%%%%%%%%%%%%%%%%%%%%%%%%%%%%%%%%%%%%%%%%%%%%

Despite its simplicity, the Shockley model may be directly relevant to the description of real materials, such as Bi$_2$Se$_3$. The crystal of Bi$_2$Se$_3$ is formed by a sequence of quintuple layers \cite{H-Zhang-2009,Liu-2010,Fu-2011,Analytis-2010,Hor-2010}. Each quintuplet consists of five alternating layers of Bi and Se, as sketched in Fig.~\ref{fig:cuts}(a). Chemical bonding within the quintuplets is relatively strong, whereas the inter-quintuplet Van der Waals attraction is relatively weak. So, the natural cleavage plane lies between the quintuplets, as shown in Fig.~\ref{fig:cuts}(a).

For the relevant energy interval near the Fermi level, the electronic structure can be captured by considering the electronic orbitals localized near the outermost layers of Se within the quintuplets \cite{Fu-2011}, as shown by the thick lines in Fig.~\ref{fig:cuts}(a). Then, the Shockley amplitudes $t_1$ and $t_2$ describe the intra- and inter-quintuplet tunneling between these orbitals, as shown in Fig.~\ref{fig:cuts}(a). The tunneling amplitudes $t_1$ and $t_2$ may depend on the in-plane momentum $\bm p$.

As shown in the previous section, the Shockley surface states strongly depend on how the crystal is terminated. When the crystal is cut between the quintuplets, and $t_2$ is broken on the surface as shown Fig.~\ref{fig:cuts}(a) and realized experimentally, a single Dirac cone is observed at the BZ center~\cite{DHsieh-2009}, as shown in Fig.~\ref{fig:cuts}(c). So, in terms of the Shockley model, the surface state existence criterion $|t_1(\bm p)|<|t_2(\bm p)|$ is satisfied at the BZ center and not satisfied at the BZ boundary.

In principle, the surface can also be introduced by cutting the quintuplet layer and breaking the bond $t_1$, as shown in Fig.~\ref{fig:cuts}(b). To the best of our knowledge, this type of surface has not been observed in Bi$_2$Se$_3$. In the previous section, we found that, for alternative crystal terminations, the surface states have the Dirac cones at the complementary TRIM points of the 2D BZ. Thus, we conclude that, when the quintuplet is broken at the surface as in Fig.~\ref{fig:cuts}(b), the surface states should have Dirac cones at the boundary of the 2D  BZ as shown in Fig.~\ref{fig:cuts}(d). A similar prediction was made for the Bi$_x$Sb$_{1-x}$ alloy in Ref.~\cite{Teo-2008}.

We can estimate the Shockley tunneling amplitudes $t_1(\bm p)$ and $t_2(\bm p)$ for $\bm p$ close to the $\Gamma$ point  based on the band-structure calculations of Ref.~\cite{H-Zhang-2009}. As discussed in Sec.~\ref{sec:classicalSchockley}, the extreme values of the energy gap can be obtained from the off-diagonal matrix element $\left.t(k)=t_1+t_2e^{ik}\right|_{k=0,\pi}=t_1\pm t_2$. We compare these values with the band structure along the direction $k\in(0,\pi)$ for the fixed in-plane momentum $\bm p=0$, which is shown in Fig.~2(b) of Ref.~\cite{H-Zhang-2009}. From the set of equations $t_1+t_2 = 0.28$~eV and $t_1-t_2 =-0.6$~eV, we obtain the following estimate
\begin{eqnarray}
 t_1 = -0.16\,{\rm eV},\quad t_2 = 0.44\,{\rm eV} \label{at1}.
\end{eqnarray}
Note that the geometric distance between the orbitals on the adjacent quintuplets is shorter than the distance between the orbitals within the quintuplet. Therefore, in the vicinity of the $\Gamma$ point, the inter-quintuplet tunneling $|t_2|$ should be greater than the intra-quintuplet tunneling $|t_1|$, which is consistent with Eq.~(\ref{at1}).

%%%%%%%%%%%%%%%%%%%%%%%%%%%%%%%%%%%%%%%%%%%%%%%%%%%%%%%%%%%%%%%%%%%%%%%%%%%%%
\subsection{Continuous approximation} \label{sec:continuous}
%%%%%%%%%%%%%%%%%%%%%%%%%%%%%%%%%%%%%%%%%%%%%%%%%%%%%%%%%%%%%%%%%%%%%%%%%%%%%

In previous sections, we have shown that, in the Shockley model, existence of the surface states relies explicitly on the tight-binding nature of the model. However, continuous models \cite{H-Zhang-2009,Liu-2010,Linder-2009,Shen-2010,Shan-2010a,Lu-2010} are also widely used to describe the surface states in the TI models and in real materials, such as Bi$_2$Se$_3$. A continuous approximation is obtained by expanding the Hamiltonian in the powers of the momentum $k$ in the $z$ direction. This is equivalent to disregarding the BZ periodicity for the momentum $k$ and taking the limit where the size of the elementary cell in the $z$ direction goes to zero. In this subsection, we examine the applicability of the continuous-limit approximation.

%%%%%%%%%%%%%%%%%%%%%%%%%%%%%%%%%%%%%%%%%%%%%%%%%%%%%%%%%%%%%%%%%%%%%%%%%%%%%
\subsubsection{First-order expansion} \label{sec:firstOrder}
%%%%%%%%%%%%%%%%%%%%%%%%%%%%%%%%%%%%%%%%%%%%%%%%%%%%%%%%%%%%%%%%%%%%%%%%%%%%%
Let us consider the Shockley Hamiltonian~(\ref{BlochHam5}) for the fixed value of the in-plane momentum $\bm p=\bm p_\ast$, where the diagonal elements vanish
\begin{eqnarray}
&&  H(k) = \left(%
\begin{array}{cc}
  0 & t^\ast(k) \\
  t(k) & 0 \\
\end{array}%
\right), \label{hmm}\\
&&  t(k) = t_1+t_2e^{ik}. \label{tkor}
\end{eqnarray}
Without loss of generality, let us make an assumption that $t_2>0$. If the energy gap $|t(k)|$ reaches minimum at $k=0$, then $t_1<0$ as in Eq.~(\ref{at1}). Then, we expand $t(k)$ to the first order in $k$ around $k=0$
\begin{equation}
  t(k) = t_1+t_2+it_2k. \label{tkkk}
\end{equation}
The tight-binding boundary conditions~(\ref{zbc}) and (\ref{ibc}) correspond the following boundary conditions~\cite{Fu-2011} for the continuous approximation
\begin{eqnarray}
&& \psi_a(z\rightarrow 0) \neq 0,\quad \psi_b(z\rightarrow 0) = 0, \label{cbc} \\
&& \psi_{a,b}(z\rightarrow\infty) = 0.\label{cibc}
\end{eqnarray}
Using these boundary conditions, we solve the Schr\"odinger equation $H\Psi=0$ for Hamiltonian~(\ref{hmm}) with the continuous $t(k)$, Eq.~(\ref{tkkk}), and obtain the surface state
\begin{equation}
  \Psi_0(z) = \left(%
\begin{array}{c}
  1 \\
  0 \\
\end{array}%
\right) e^{ik_0z}. \label{psi2}
\end{equation}
The exponential decay length in Eq.~(\ref{psi2}) is given by the parameter
\begin{equation}
  k_0 = i\left(1+\frac{t_1}{t_2}\right),  \label{rootApprox}
\end{equation}
which is the root of the equation $t(k_0)=t_1+t_2+it_2k_0=0$. Boundary conditions~(\ref{cibc}) are satisfied if Im~$k_0>0$ or equivalently $t_1>-t_2$; otherwise, the surface state does not exist if $t_1<-t_2$. So, the continuous model~(\ref{tkkk}) with the appropriate boundary conditions~(\ref{cbc}) and (\ref{cibc}) gives the surface state existence criterion
\begin{equation}
 |t_1|<|t_2|,
\end{equation}
which coincides with the Shockley criterion~(\ref{crit}). The continuous wave function~(\ref{psi2}) correctly approximates the discrete wave function~(\ref{psi}) if the decay length is very long or equivalently $||t_1|-|t_2||\ll |t_2|$. However, the estimated tunneling amplitudes $t_1$ and $t_2$ in Eq.~(\ref{at1}) do not satisfy the latter condition for Bi$_2$Se$_3$. Therefore, we conclude that the discrete Shockley model gives a more appropriate  description  of the surface states in ${\rm Bi}_2{\rm Se}_3$, than a continuous approximation, because the difference between $|t_1|$ and $|t_2|$ is rather large.

%%%%%%%%%%%%%%%%%%%%%%%%%%%%%%%%%%%%%%%%%%%%%%%%%%%%%%%%%%%%%%%%%%%%%%%%%%%%%
\subsubsection{Higher-order expansion} \label{sec:higherOrder}
%%%%%%%%%%%%%%%%%%%%%%%%%%%%%%%%%%%%%%%%%%%%%%%%%%%%%%%%%%%%%%%%%%%%%%%%%%%%%

%%%%%%%%%%%%%%%%%%%%%%%%%%%%%%%%%%%%%%%%%%%%%%%%%%%%%%%%%%%%%%%%%%%%%%%%%%%%%
\begin{figure}
 \includegraphics[width=\linewidth]{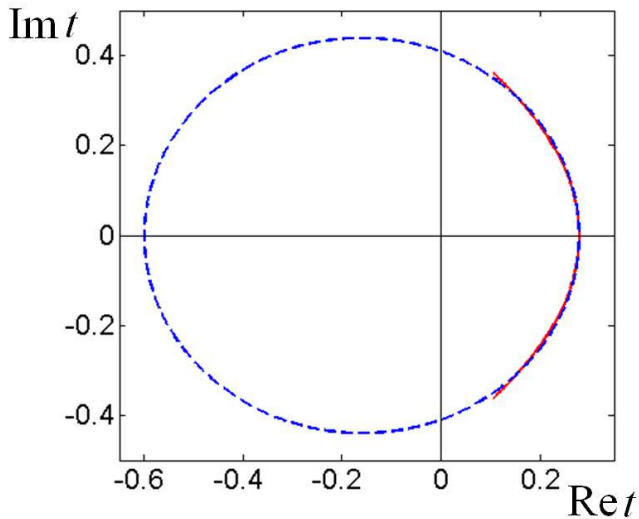}
 \caption{(Color online) Plot of the function $t(k)$ in the complex plane of $({\rm Re}\,t,{\rm Im}\,t)$. The second-order expansion for $t(k)$, given by Eq.~(\ref{conttk}) with the parameters from Ref.~\cite{Liu-2010}, is plotted by the solid line for $-\pi/2<k<\pi/2$. The function $t(k)$ in the Shockley model, given by Eq.~(\ref{tkor}) with the parameters $t_1$ and $t_2$ from Eq.~(\ref{at1}), is plotted by the dashed line. The Shockley contour winds around the origin, which guarantees existence of the surface state.} \label{fig:windingContoursComparison}
\end{figure}
%%%%%%%%%%%%%%%%%%%%%%%%%%%%%%%%%%%%%%%%%%%%%%%%%%%%%%%%%%%%%%%%%%%%%%%%%%%%%

One may truncate the series for $e^{ik}$ in Eq.~(\ref{tkkk}) at a higher order in $k$
\begin{equation}
  t(k) = t_1+t_2+t_2\sum_{n=1}^N \frac{(ik)^n}{n!}. \label{eqmom}
\end{equation}
However, such a truncation gives worse continuous description of the Shockley surface state.  The equation $t(k)=0$ now has $N$ roots $k_1,\ldots ,\,k_N$. So, there are $N$ independent coefficients $c_n$ in a general solution $\Psi(z)=c_1e^{ik_1z}+\ldots +c_Ne^{ik_Nz}$ to satisfy the boundary conditions (\ref{cbc}) and (\ref{cibc}). This gives rise to  a  large number of the unphysical surface state solutions, while the Shockley model predicts only one surface state.  Most of the roots $k_j$ have large imaginary parts Im~$k_j\gtrsim1$. These solutions are spurious, because they correspond to the wave functions decaying over a length shorter than the unit cell of the crystal. For example, for $N=2$, Eq.~(\ref{eqmom}) is
\begin{equation}
  t(k) = t_1+t_2+it_2k-t_2^2k^2/2. \label{eqmom1}
\end{equation}
Then, the equation $t(k)=0$ has two roots
\begin{equation}
  k_{1,2} = i\pm\sqrt{-1+2(1+t_1/t_2)}. \label{solroots}
\end{equation}
In the limit $|1+t_1/t_2|\ll 1$, the roots become $k_1=i(1+t_1/t_2)$ and $k_2 = 2i$. We observe that, while the first root $k_1$ reproduces the correct approximation Eq.~(\ref{rootApprox}), the second root $k_2$ has a large imaginary part and must be discarded.  In another regime, when the expression under the square root in Eq.~(\ref{solroots}) is positive, both roots have large imaginary parts Im~$k_{1,2} = 1$, so the continuous approximation is not applicable.  Moreover, as pointed out in Ref.~\cite{Fu-2011}, the continuous description does not distinguish between two possible ways of terminating the crystal shown in Fig.~\ref{fig:cuts}(a) and (b). A correct boundary condition should be chosen to distinguish between different possible surface terminations.

Despite these problems, the $k^2$ terms were kept in the effective description of Bi$_2$Se$_3$ in Ref.~\cite{Liu-2010}
\begin{eqnarray}
&&  H = H_0+H_1, \label{continn}\\
&&  H_0 = \epsilon(k)+(M_0+M_1k^2)\tau_z+B_0k\tau_y, \label{H000} \\
&&  H_1 = A_0 \tau_x (\bm\sigma\times\bm p), \label{H111}
\end{eqnarray}
where $ M_0 = -0.28\,{\rm eV},\,\, M_1 = 6.86\,{\rm eV \text{\AA}^2},\,\,B_0 =2.26\,{\rm eV\text{\AA}},\,\,A_0 =3.33\,{\rm eV\text{\AA}}$; $\tau_y$ and $\tau_x$ are the Pauli matrices. In Eq.~(\ref{H111}), $H_1$ represents spin-orbit interaction and explicitly depends on the spin operators $\bm \sigma$ and the in-plane momentum $\bm p$. $H_0$ depends on the out-of-plane momentum $k$ and is responsible for the existence of the surface states. Following Ref.~\cite{Liu-2010}, we drop the term $\epsilon(k)$ in Eq.~(\ref{H000}), because it is proportional to the unit matrix. Then we apply the unitary transformation $e^{-i\tau_y\pi/4}$, which changes $\tau_z\rightarrow-\tau_x$ and $\tau_x\rightarrow\tau_z$. So, the Hamiltonian becomes $U^\dag H_0U\rightarrow H_0$
\begin{eqnarray}
  && H_0 =\left(
         \begin{array}{cc}
           h(\bm p) & t^\ast(k) \\
           t(k)& -h(\bm p) \\
         \end{array}
       \right), \label{hamm3} \\
  && t(k) = -M_0-M_1k^2+ iB_0k,  \label{conttk} \\
  && h(\bm p) = A_0 (\bm\sigma\times\bm p). \label{hmp}
\end{eqnarray}
Now, the Hamiltonian $H_0$ has the same form as in Eq.~(\ref{BlochHam5}). Following Ref.~\cite{Fu-2011}, we infer that the basis for the Hamiltonian~(\ref{hamm3}) corresponds to the basis of electron orbitals located at the outermost layers of the quintuplet. Then, the off-diagonal matrix element $t(k)$ in Eq.~(\ref{conttk}) corresponds to the second-order expansion of the off-diagonal element of the Shockley model~(\ref{eqmom1}), while $h(\bm p)$ defines the in-plane dispersion.

To make explicit correspondence with the previous section, we change units for $k$: $k a\rightarrow k$, where $a=1$~nm is the size of the elementary cell of Bi$_2$Se$_3$ in the $z$ direction. So, we rewrite the parameters $M_1/a^2\rightarrow M_1$, $B_0/a\rightarrow B_0$ in the energy units of eV
\begin{eqnarray}
  M_0 = -0.28\,{\rm eV}, \,\, M_1 = 0.07\,{\rm eV},\,\,  B_0 = 0.23\,{\rm eV}. \nonumber
\end{eqnarray}
Notice that $t(k)$, Eq.~(\ref{conttk}), parametrizes a parabola in the complex space $($Re $t$,Im $t)$ when $k$ is changed. So we plot $t(k)$ defined by Eq.~(\ref{conttk}) for $-\pi/2<k<\pi/2$ by the solid line in Fig.~\ref{fig:windingContoursComparison}. Figure~\ref{fig:windingContoursComparison} also shows the plot $t(k)$  for the discrete Shockley model~(\ref{hmm}) with the parameters~(\ref{at1}) by the dashed line.  We see that the continuous approximation to the Hamiltonian agrees with the Shockley model within a limited range of $k$ with the continuous approximation. Nevertheless, the continuous approximation has serious deficiencies for construction of the wave functions, as described above, whereas the Shockley model provides a good overall description for the surface states in Bi$_2$Se$_3$.

%%%%%%%%%%%%%%%%%%%%%%%%%%%%%%%%%%%%%%%%%%%%%%%%%%%%%%%%%%%%%%%%%%%%%%%%%%%%%
\section{Generalized Shockley model} \label{sec:generalizedShockley}
%%%%%%%%%%%%%%%%%%%%%%%%%%%%%%%%%%%%%%%%%%%%%%%%%%%%%%%%%%%%%%%%%%%%%%%%%%%%%
In this section, we generalize the Shockley model to include additional inter-cell tunneling amplitudes. To simplify notations, we present results for the 1D case. However, the results can be straightforwardly  generalized to the 3D case by assigning dependence on $\bm p=(p_x,p_y)$ to the tunneling amplitudes, as discussed in Sec.~\ref{sec:Shockley}.
%%%%%%%%%%%%%%%%%%%%%%%%%%%%%%%%%%%%%%%%%%%%%%%%%%%%%%%%%%%%%%%%%%%%%%%%%%%%%
\subsection{Additional tight binding amplitude \boldmath$t_3$} \label{sec:t3}
%%%%%%%%%%%%%%%%%%%%%%%%%%%%%%%%%%%%%%%%%%%%%%%%%%%%%%%%%%%%%%%%%%%%%%%%%%%%%

%%%%%%%%%%%%%%%%%%%%%%%%%%%%%%%%%%%%%%%%%%%%%%%%%%%%%%%%%%%%%%%%%%%%%%%%%%%%%
\begin{figure}
 \includegraphics[width=\linewidth]{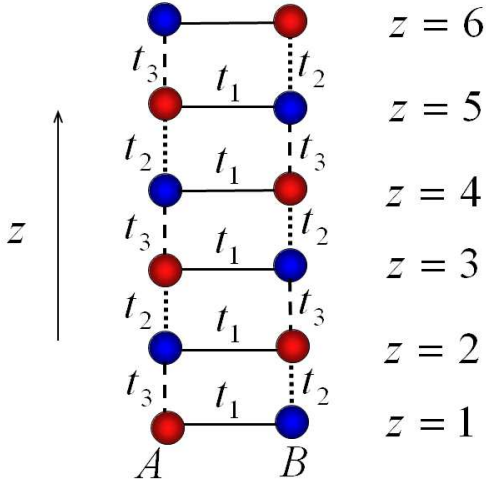}
 \caption{(Color online)  An illustration of the generalized Hamiltonian~(\ref{BlochHam2}). The Hamiltonian describes a tight-binding model with the elementary cell comprised of the A and B sites, which are connected via the complex tight-binding amplitudes $t_1$, $t_2$ and $t_3$.} \label{fig:Shockleya}
\end{figure}
%%%%%%%%%%%%%%%%%%%%%%%%%%%%%%%%%%%%%%%%%%%%%%%%%%%%%%%%%%%%%%%%%%%%%%%%%%%%%

Let us consider a 1D Hamiltonian of the form given by Eq.~(\ref{Ham}) with
\begin{equation}
   U = \left(%
\begin{array}{cc}
  0 &  t_1^\ast \\
  t_1 & 0 \\
\end{array}%
\right),\,\,\,\,
V = \left(%
\begin{array}{cc}
  0 &  t_2^\ast \\
  t_3 & 0 \\
\end{array}%
\right), \label{uv1}
 \end{equation}
 where the matrix $V$ now contains an additional tight-binding amplitude $t_3$. The 1D chain model corresponding to Eq.~(\ref{uv1}) is illustrated in Fig.~\ref{fig:Shockleya}.  The amplitude $t_1$ describes tunneling between the A and B sublattices inside the unit cell, and the amplitudes $t_2$ and $t_3$ between the unit cells. The introduction of this tight-binding amplitude is motivated by the TI literature \cite{Hosur-2010,Mong-2011} as well by the novel 1D models such as the superconducting Majorana chain \cite{Kitaev-2001,Bermudez-2010} and the Creutz ladder \cite{Creutz-1994,Bermudez-2009}. This model is a natural mathematical generalization of the models considered in the previous sections. The Hamiltonian of the general model has the same form as in Eq.~(\ref{BlochHam}),
\begin{equation}
  H(k) = \left(%
\begin{array}{cc}
  0 & t^\ast(k) \\
  t(k) & 0 \\
\end{array}%
\right), \label{BlochHam2}
\end{equation}
with
\begin{equation}
  t(k)=t_1+t_2e^{ik}+t_3e^{-ik}.
\end{equation}

\begin{figure}
 \includegraphics[width=\linewidth]{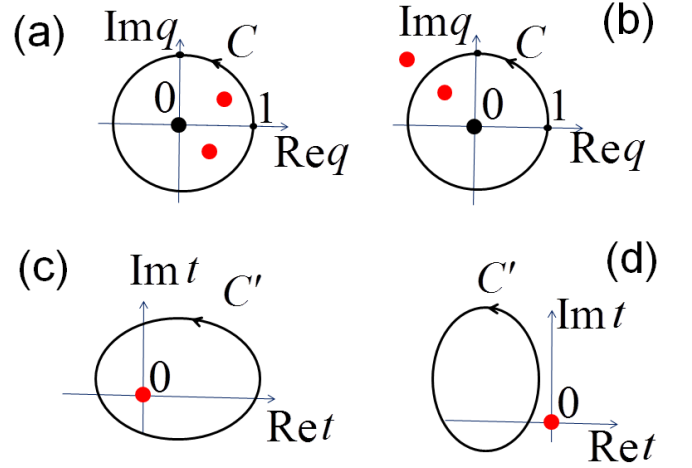}
\caption{(Color online)  Topological formulation of the Shockley criterion. The roots $q_{1,2}$ of Eq.~(\ref{ImEq}) are shown in panels (a) and (b) by red dots. An edge state exists if the roots are on the same side of the unit circle $C=\{q=e^{ik},\,k\in(0,2\pi)\}$, as shown in panel (a). No edge state exists if the roots are on the opposite sides of the unit circle, as shown in panel (b). The thick black dot at the origin is the pole of Eq.~(\ref{ImEq}). An alternative formulation in terms of the winding number~(\ref{critWinding1}) is shown in panels (c) and (d). An edge state exists if the contour $C'=\{t(k),\,k\in(0,2\pi)\}$ winds around the origin, as shown in panel (c). No edge state exists if the contour $C'$ does not wind around the origin, as shown in panel (d). }
\label{fig:windingContoursG}
\end{figure}

As in Eqs.~(\ref{A}) and (\ref{B}), the eigenstate equations for the wave functions on the $A$ and $B$ sublattices decouple at $E=0$. (Appendix~\ref{sec:appendixB} proves that the edge state can exist only for $E=0$.) The zero-energy state on the A sublattice has the complex momentum $k$ obtained from the equation
\begin{eqnarray}
  t(k)=t_1+t_2e^{ik}+t_3e^{-ik}=0. \label{eqq1}
\end{eqnarray}
We substitute $q = e^{ik}$ and obtain an equation for the rational function $t(q)$
\begin{eqnarray}
  t(q)=t_1+t_2q+t_3q^{-1}=0 \label{ImEq},
\end{eqnarray}
which has two solutions
\begin{equation}
  q_{1,2}=e^{ik_{1,2}}=\frac{1}{2t_2}\left
  (-t_1\pm\sqrt{t_1^2-4t_2t_3}\right), \label{ssol}
\end{equation}
with the complex momenta $k_{1,2}$. Using these momenta, we construct an edge state that satisfies the boundary conditions given by Eqs.~(\ref{zbc}) and (\ref{ibc}). The edge state has the energy $E_0=0$ and is localized on the A sublattice
\begin{eqnarray}
&   \Psi_0(z) = \left(%
\begin{array}{c}
  \psi_a^{(0)}(z) \\
  0 \\
\end{array}%
\right), \quad E_0 =0,\\
&   \psi_a^{(0)}(z)= q_1^z-q_2^z = e^{ik_1z}-e^{ik_2z}.
\label{psia}
\end{eqnarray}
The wave function~(\ref{psia}) satisfies the  boundary condition (\ref{ibc}) if Im~$k_1>0$ and Im~$k_2>0$ or, equivalently,
\begin{eqnarray}
  |q_1|<1{\rm \,\, and\,\,}|q_2|<1. \label{ccra}
\end{eqnarray}
Likewise, a zero-energy state on the B sublattice has the complex momenta $k$ obtained from the equation
\begin{equation}
  t^\ast(k)=t_1^\ast+t_2^\ast e^{-ik}+t_3^\ast e^{ik}=0.\label{eqq2}
\end{equation}
Notice that the symbol of complex conjugation $\ast$ applies only to the tunneling amplitudes in Eq.~(\ref{eqq2}), so $t^\ast(k)\neq [t(k)]^\ast$ if ${\rm Im}\,k \neq 0$. Equation~(\ref{eqq2}) can be obtained by replacing $k\rightarrow k^\ast$ in Eq.~(\ref{eqq1}). So, the two solutions $k'_{1,2}$ of Eq.~(\ref{eqq2}) and the corresponding $q'_{1,2}$ can be obtained from Eq.~(\ref{ssol})
\begin{equation}
  k'_{1,2}=k_{1,2}^\ast,\quad q'_{1,2}=1/q_{1,2}^\ast.
\end{equation}
The edge state exists on the B sublattice
\begin{eqnarray}
&   \Psi_0(z) = \left(%
\begin{array}{c}
  0 \\
  \psi_b^{(0)}(z) \\
\end{array}%
\right),\,\,E_0=0, \\
&   \psi_b^{(0)}(z)= (q'_1)^z-(q'_2)^z=e^{ik'_1z}-e^{ik'_2z}
\label{psib}
\end{eqnarray}
if Im~$k'_1>0$ and Im~$k'_2>0$ or, equivalently,
\begin{eqnarray}
  |q_1|>1\,\,{\rm and}\,\,|q_2|>1. \label{ccrb}
\end{eqnarray}
To summarize, the edge state~(\ref{psia}) exists on the A sublattice if both roots of Eq.~(\ref{ImEq}) are inside the unit circle, as in Eq.~(\ref{ccra}) and in Fig.~\ref{fig:windingContoursG}(a). The edge state~(\ref{psib}) exists on the B sublattice if both roots of Eq.~(\ref{ImEq}) are outside the unit circle, as in Eq.~(\ref{ccrb}). The edge state does not exist if one of the roots is inside and another root is outside the unit circle
\begin{equation}
  |q_1|>1\,\,{\rm and}\,\,|q_2|<1, \label{ccrd}
\end{equation}
as shown in Fig.~\ref{fig:windingContoursG}(b).  Obviously, the conditions~(\ref{ccra})~and~(\ref{ccrb}) cannot be met simultaneously, so edge state cannot exist on both sublattices simultaneously.

Like in Sec.~\ref{sec:classicalSchockley}, the criterion for the edge states existence can be formulated in terms of the winding number of the complex function $t(q)$ along the unit circle  $C=\{|q|=1\}$
\begin{equation}
  W = \frac{1}{2\pi i}\oint\limits_{|q|=1} dq\,\frac{d}{dq}{\rm ln}\,\left[t(q)\right].
 \label{winding1}
\end{equation}
The criteria given by Eqs.~(\ref{ccra}),~(\ref{ccrb}),~and~(\ref{ccrd}) are summarized in the following
\begin{equation}
 W = \left\{\begin{array}{c}
   1, {\rm \,\,edge\,\,state\,\,}\psi_a^{(0)}(z)\,\,{\rm exists},  \\
   0, {\rm \,\,edge\,\,state\,\,does\,\,not\,\,exist}, \\
   -1, {\rm \,\,edge\,\,state\,\,}\psi_b^{(0)}(z)\,\,{\rm exists}. \\
 \end{array}\right. \label{critWinding1}
\end{equation}
To prove it, we use Cauchy's argument principle
\begin{equation}
 W=Z-P, \label{Cauchy}
\end{equation}
which relates the winding number $W$ of a complex function $t(q)$ on a contour $C$ with the number of zeros $Z$ and the number of poles $P$ inside the contour $C$. Since $t(q)$ given by Eq.~(\ref{ImEq}) has a pole at $q=0$, as shown by the thick black dot in Fig.~\ref{fig:windingContoursG}(a) and (b), the number of poles is $P=1$. The edge state exists on the A sublattice if $|q_{1,2}|<1$, in which case $W = Z-P=2-1=1$. The edge state exists on the B sublattice if $|q_{1,2}|>1$, in which case $W=Z-P=0-1=-1$. The edge state does not  exist for $|q_1|>1$ and $|q_2|<1$, in which case $W=Z-P=1-1=0$.

In other words, according to Eq.~(\ref{critWinding1}), the edge state exists if the closed contour
\begin{equation}
  C' = \{t(k),\,k\in(0,\,2\pi)\} \label{curve}
\end{equation}
winds around the origin, as shown in Fig.~\ref{fig:windingContoursG}(c). The direction of winding of $t(k)$ defines the sublattice on which the edge state is localized. An analogous criterion was proposed in Ref.~\cite{Mong-2011} (for a comparison with our model, see Appendix~\ref{sec:appendixC}).

For the tunneling amplitudes $t_1$, $t_2$, and $t_3$ connecting the nearest unit cells, Eq.~(\ref{curve}) defines an ellipse
\begin{equation}
  t(k)=t_1+(t_2+t_3)\cos k+i(t_2-t_3)\sin k, \label{ellipse}
\end{equation}
which is  shifted by $t_1$ from the origin. In case where the tunneling amplitudes are real, the ellipse in Eq.~(\ref{ellipse}) encloses the origin if
\begin{equation}
|t_1|<|t_2+t_3|. \label{ShockleyGeneral}
\end{equation}
Eq.~(\ref{ShockleyGeneral}) represents the generalized Shockley rule of a stronger bond: The edge state exists if the broken inter-cell bond $t_2+t_3$ is stronger than the intra-cell bond $t_1$.

Our consideration does not include tunneling amplitudes connecting sites on the same sublattices in different unit cells. Including such terms would make $h$ in Eq.~(\ref{BlochHam5}) depend on $k$. When these tunneling amplitudes connect only the nearest neighboring unit cells, the problem can still be solved as shown in Ref.~\cite{Mong-2011} (see a discussion in Appendix~\ref{sec:appendixC}). 
%%%%%%%%%%%%%%%%%%%%%%%%%%%%%%%%%%%%%%%%%%%%%%%%%%%%%%%%%%%%%%%%%%%%%%%%%%%%%
%%%%%%%%%%%%%%%%%%%%%%%%%%%%%%%%%%%%%%%%%%%%%%%%%%%%%%%%%%%%%%%%%%%%%%%%%%%%%

%%%%%%%%%%%%%%%%%%%%%%%%%%%%%%%%%%%%%%%%%%%%%%%%%%%%%%%%%%%%%%%%%%%%%%%%%%%%%
\subsection{Arbitrary periodic function \boldmath$t(k)$} \label{sec:arbitrary}
%%%%%%%%%%%%%%%%%%%%%%%%%%%%%%%%%%%%%%%%%%%%%%%%%%%%%%%%%%%%%%%%%%%%%%%%%%%%%

In the most general form, Hamiltonian~(\ref{BlochHam2}) can be written as
\begin{equation}
  H(k) = \left(%
\begin{array}{cc}
  0 & t^\ast(k) \\
  t(k) & 0 \\
\end{array}%
\right), \label{BlochHam9}
\end{equation}
where
\begin{equation}
  t(k)=\sum_{n=-N}^N t_ne^{ikn}. \label{tkgen}
\end{equation}
This model describes a 1D tight-binding chain, where each unit cell is coupled to  $N$ preceding and $N$ successive unit cells. Therefore, for a half-infinite system at $z\ge 1$, the boundary conditions require that the wave function vanishes at the fictitious $N$ sites adjacent to the boundary
\begin{eqnarray}
  \Psi(-N+1) = 0,\,\,\,\,\ldots,\,\,\,\,\Psi(-1)=0,\,\,\Psi(0) = 0, \label{bouncond}
\end{eqnarray}
similarly to Eq.~(\ref{zbc}). As in the previous section, let us substitute $q=e^{ik}$ and rewrite Eq.~(\ref{tkgen}) in the polynomial form
\begin{equation}
 t(q)=\sum_{n=-N}^N t_nq^n. \label{tkpolynome}
\end{equation}
This polynomial has $2N$ roots. Suppose, the number of roots $N_1$ with $|q_j|<1$, $j=1,\cdots,N_1$, is greater than $N$: $N_1>N$. In this case, we can construct a trial wave function
\begin{equation}
  \Psi(z) = \sum_{j=1}^{N_1} \alpha_j q_j^z,
\end{equation}
which vanishes at $z\rightarrow\infty$. The coefficients $\alpha_j$ in Eq.~(\ref{bouncond}) are obtained by solving a set of $N$ boundary condition equations (\ref{bouncond}). Therefore, in general, there are $N_1-N$ linearly independent solutions for the edge states localized on the sublattice A.  The same result can be formulated using the winding number in Eq.~(\ref{winding1}). Indeed, the function~(\ref{tkpolynome}) has a pole of the $N$-th order at $q=0$ and $N_1$ zeros at $|q_j|<1, j=1,\ldots,N_1$. Therefore, using Cauchy's argument principle~(\ref{Cauchy}), we obtain
 \begin{equation}
 W = Z-P = N_1-N.
\end{equation}
Thus, the winding number $W$ of the function $t(k)$ gives the number of the edge states. On the other hand, if $W<0$, then there are $|W|$ degenerate edge states localized on the sublattice B. The edge states cannon exist simultaneously on both sublattices A and B. There are no edge states for $W=0$. Finally, in the limit $N\rightarrow\infty$, the winding criterion applies to an arbitrary complex function $t(k)$ periodic in $k$.

%%%%%%%%%%%%%%%%%%%%%%%%%%%%%%%%%%%%%%%%%%%%%%%%%%%%%%%%%%%%%%%%%%%%%%%%%%%%%
\section{Symmetries} \label{sec:symmetries}
%%%%%%%%%%%%%%%%%%%%%%%%%%%%%%%%%%%%%%%%%%%%%%%%%%%%%%%%%%%%%%%%%%%%%%%%%%%%%

In this section, we discuss the symmetries of the Shockley model. Let us first consider the case $h(\bm p)=0$ in the generalized Shockley Hamiltonian~(\ref{BlochHam5})
\begin{equation}
  H(k,\bm p) = \left(%
\begin{array}{cc}
  0 & t^\ast(k,\bm p) \\
  t(k,\bm p) & 0 \\
\end{array}%
\right). \label{BlochHam10}
\end{equation}
In this case, the A and B sublattices have equal on-site energies, which is reflected by a chiral symmetry of the Hamiltonian: $\tau_z H(k,\bm p) \tau_z = -H(k,\bm p)$, where $\bm \tau = (\tau_x,\tau_y,\tau_z)$ are the Pauli matrices acting in the AB sublattice space.  Therefore, Hamiltonian~(\ref{BlochHam10}) belongs to the class AIII of chiral Hamiltonians \cite{Schnyder-2008}. As a consequence of chiral symmetry, the energy spectrum is symmetric: if $\Psi$ is an eigenstate $H\Psi = E\Psi$, then $\tau_z\Psi$ is also an eigenstate corresponding to the opposite energy $H\tau_z\Psi = -E \tau_z\Psi$. Therefore, if a non-degenerate eigenstate with $E=0$ exists, it should be an eigenstate of $\tau_z\Psi = \lambda\Psi$, $\lambda=\pm 1$. So, the $E=0$ state must be localized on one of the sublattices, consistently with Eq.~(\ref{psi}). The winding number $W\in Z$ of the vector $\bm d (k,\bm p)$, defined in Eq.~(\ref{TauRepresent}), gives the number of surface states for a fixed $\bm p$.

In Sec.~\ref{sec:shockleyOnSiteEnergy}, we generalized the Shockley model by including the diagonal element $h(\bm p)$ 
\begin{equation}
  H(k,\bm p) = \left(%
\begin{array}{cc}
  h(\bm p) & t^\ast(k,\bm p) \\
  t(k,\bm p) & -h(\bm p) \\
\end{array}%
\right). \label{BlochHam11}
\end{equation}
The Hamiltonian~(\ref{BlochHam11}) does not have a chiral symmetry, but it has another sublattice symmetry ($i\tau_yK)\,H\,(i\tau_yK) = H$, where $K$ is the operator of complex conjugation. This symmetry exchanges the sublattices, $i\tau_y K\, (\psi_A\,,\, \psi_B)^T =(\psi_B^\ast\,,\,-\psi_A^\ast)^T$, and makes the bulk spectrum symmetric (there is an opposite energy counterpart $i\tau_y K \Psi$ for every eigenstate $\Psi$). However, this symmetry is broken at the boundary of the crystal, where one of the sublattices is exposed at the surface, as shown in Fig.~\ref{fig:layers}. As a result, the surface state gains the dispersion $E_0(\bm p) = h(\bm p)$, whereas the opposite-energy counterpart of the surface state is localized at the opposite surface.  The number of the surface states is still given by the winding number $W\in Z$ of the vector $\bm d (k,\bm p)$ as a function $k$ for a fixed $\bm p$.

In this work, we discussed Hamiltonians with the time-reversal symmetry in the presence of spin-orbit coupling, which belong to the class AII of topological insulators classification \cite{Schnyder-2008}. The spin-orbit coupling and the time-reversal symmetry require the term $h(\bm p)$ to have the 2D Dirac-type form, as discussed in Sec.~\ref{sec:SpinOrbit}. However, the Shockley Hamiltonian~(\ref{BlochHam11}) is applicable in a more general case, where the term $h(\bm p)$ is an arbitrary Hermitian matrix not necessarily respecting the time-reversal symmetry. So, the Shockley Hamiltonian~(\ref{BlochHam11}) can describe the quantum Hall states, which belong to the class A of topological insulators classification \cite{Schnyder-2008}. In addition, the Shockley model can describe superconducting systems, in which case Eq.~(\ref{BlochHam11}) should be understood as a Bogolyubov-de Gennes Hamiltonian~\cite{Wimmer-2010,Beenakker-2011,Beri-2010,Schnyder-2010}, and $i\tau_yK$ represents the particle-hole symmetry.

%%%%%%%%%%%%%%%%%%%%%%%%%%%%%%%%%%%%%%%%%%%%%%%%%%%%%%%%%%%%%%%%%%%%%%%%%%%%%
\section{Conclusions} \label{sec:conclusion}
%%%%%%%%%%%%%%%%%%%%%%%%%%%%%%%%%%%%%%%%%%%%%%%%%%%%%%%%%%%%%%%%%%%%%%%%%%%%%
Like some previous works \cite{Mong-2011,Heikkila-2011,Hosur-2010},  our paper explores a tight-binding theory of the surface states in topological insulators. We show that the surface states can be understood using the simple and well-known Shockley model \cite{Shockley-1939,Davison-1996}, a 1D model with the A and B atoms per unit cell, connected via alternating tunneling amplitudes. We generalize the 1D Shockley model to the 3D case described by the $2\times 2$ Hamiltonian~(\ref{BlochHam5}) with the diagonal element $h(\bm p)$ and the off-diagonal element $t(k,\bm p)$. The diagonal element $h(\bm p)$ defines the energy dispersion of the surface states, while the complex-valued off-diagonal element $t(k,\bm p)$ defines the domain of existence of the surface states. The surface states exist for those in-plane momenta $\bm p$ where the phase winding of $t(k,\bm p)$ along $k\in(0,2\pi)$ is non-zero. The sign of the winding number gives the sublattice A or B, on which the surface states are localized. Equation $t(k,\bm p)=0$ defines a vortex line in the 3D momentum space \cite{Beri-2010,Heikkila-2011,Schnyder-2010}, and projection of the vortex line onto the 2D space of $\bm p$ is the boundary of the domain where the surface states exist. We apply this approach to the TI model on the diamond lattice \cite{Kane-2007}. We show how the evolution of the vortex lines is responsible for transitions between the ``weak'' and ``strong'' TI phases. We discuss why the discrete Shockley model is better than continuous models for the description of the edge states in real materials, such as Bi$_2$Se$_3$. The tight-binding model demonstrates that different types of surface states are formed depending on how crystal is terminated~\cite{Fu-2011}. The surface states have the Dirac cone at the center of the Brillouin zone when the crystal is cut between the quintuple layers of Bi$_2$Se$_3$, but, when the crystal is terminated inside the quintuple layer, the surface states have three Dirac cones on the boundary of the Brillouin zone. We also generalize the Shockley model to an arbitrary complex function $t(k)$ periodic in $k$, which includes the long-range inter-cell tunneling amplitudes. We prove the validity of the winding number criterion in this general case as well. We hope that this work will provide a useful toolkit for studying the surface states in TI, as well as give a transparent picture for their physical interpretation.

%%%%%%%%%%%%%%%%%%%%%%%%%%%%%%%%%%%%%%%%%%%%%%%%%%%%%%%%%%%%%%%%%%%%%%%%%%%%%
\begin{acknowledgements}
 The authors thank Liang Fu for a useful discussion.
\end{acknowledgements}
%%%%%%%%%%%%%%%%%%%%%%%%%%%%%%%%%%%%%%%%%%%%%%%%%%%%%%%%%%%%%%%%%%%%%%%%%%%%%

\appendix

%%%%%%%%%%%%%%%%%%%%%%%%%%%%%%%%%%%%%%%%%%%%%%%%%%%%%%%%%%%%%%%%%%%%%%%%%%%%%
\section{Edge states in the original Shockley model} \label{sec:appendixA}
%%%%%%%%%%%%%%%%%%%%%%%%%%%%%%%%%%%%%%%%%%%%%%%%%%%%%%%%%%%%%%%%%%%%%%%%%%%%%
In Sec.~\ref{sec:Shockley}, the Schr\"odinger equation for the wave function $\Psi(z)$ was given in a recursive form for the integer coordinate $z\ge 1$. The main question is whether the recursion generates a decaying function $\Psi(z) \rightarrow  0$ at $z\rightarrow\infty$, which represents an edge state, or an increasing function $\Psi(z) \rightarrow  \infty$ at $z\rightarrow\infty$, which is unphysical. Below, we use the generating function method to find convergence criterion for the edge-state solution $\Psi(z)$. The Schr\"odinger equation for the original Shockley model~(\ref{Ham}) is
\begin{eqnarray}
& V\Psi(z)+(U-E)\Psi(z+1)+V^\dag\Psi(z+2)=0,
\label{recur}
\\
&  V = \left(%
\begin{array}{cc}
  0 & t_2 \\
  0 & 0 \\
\end{array}%
\right),\,\,\,\, U = \left(%
\begin{array}{cc}
  0 & t_1 \\
  t_1 & 0 \\
\end{array}%
\right),\label{uv}
\end{eqnarray}
where $\Psi(z)=[\psi_a(z),\psi_b(z)]^{T}$ and $z\ge 1$, whereas the boundary condition is
\begin{equation}
 (U-E)\Psi(1)+V^\dag\Psi(2)=0. \label{bc}
\end{equation}
Let us multiply the $z$-th Eq.~(\ref{recur}) by the $(z-1)$-th power of an auxiliary complex variable $q$ and take a sum for $z\ge 1$
\begin{equation}
  \sum\limits_{z=1}^\infty q^{z-1}\left [V\Psi(z)+(U-E)\Psi(z+1)+V^\dag\Psi(z+2)\right ]=0. \label{infsum}
\end{equation}
Introducing the generating function
\begin{equation}
 G(q) = \sum_{z=1}^\infty q^{z-1} \Psi(z), \label{series}
\end{equation}
Eq.~(\ref{infsum}) can be written as
\begin{equation}
 [q^2V+q(U-E)+V^\dag]G(q)=V^\dag\Psi(1),
 \label{gen0}
\end{equation}
where we utilized the boundary condition~(\ref{bc}). From Eq.~(\ref{gen0}), we obtain the generating function in terms of $\Psi(1)$
\begin{equation}
 G(q)=[q^2V+q(U-E)+V^\dag]^{-1}\,\,V^\dag\,\Psi(1),
 \label{gen}
\end{equation}
In order to investigate convergence of $\Psi(z)$, we use the following proposition

{\textbf{Proposition 1}. \it A rational generating function $G(q)$ corresponds to an edge state $\Psi(z)\xrightarrow{z\rightarrow \infty} 0$ if and only if all poles $q_{j=1,2,3,\ldots}$ of $G(q)$ have the absolute values greater than one, $|q_j|>1$.}

Indeed, a rational function with the poles $q_j$ can be transformed to the form $G(q)=\sum_j\frac{f_j(q)}{(q-q_j)^{n_j}}$, where $f_j(q)$ is a polynomial function, and $n_j$ is the order of the pole $q_j$. Consider a simple example of the first-order pole $G(q)=\frac{q_1}{q_1-q}=\sum_z (q/q_1)^z$, which corresponds to the geometric progression. According to Eq.~(\ref{series}), the expansion coefficients give the wave function $\Psi(z)=1/q_1^{z-1}$. Then, the absolute values of the pole $|q_1|<1$, $|q_1|=1$, and $|q_1|>1$ correspond, respectively, to an un-physical growing solution $\Psi(z)\xrightarrow{z\rightarrow \infty} \infty$, a bulk state $\Psi(z)=e^{ikz}$, and a decaying edge state $\Psi(z)\xrightarrow{z\rightarrow \infty} 0$. The case of a more complicated $G(q)$ can be reduced to the above simple consideration.

Now let us use Proposition~1 to investigate convergence of $\Psi(z)$.  Using Eq.~(\ref{gen}) and the expressions for $U$ and $V$ in Eq.~(\ref{uv}), we find
\begin{equation}   G(q) = \frac{\psi_a(1)\,\,t_2}{(t_1+t_2q)(t_2+t_1q)-E^2q}\left(%
\begin{array}{c}
  t_1+t_2q \\
  E q \\
\end{array}%
\right). \label{Gq}
\end{equation}
The poles of Eq.~(\ref{Gq}) are given by the zeros $q_1$ and $q_2$ of the denominator, unless they are canceled out by zeros in the numerator. Using Vieta's formulas for the quadratic equation in the denominator, we obtain $q_1q_2 = 1$. So, if $q_1$ is greater than one, $|q_1|>1$, then $q_2$ is less than one, $|q_2|<1$. Using Proposition 1, we conclude that there is no edge state if the generating function G(q) in Eq.~(\ref{Gq}) has two poles. In order to obtain an edge state, we need to reduce the number of poles of the generating function $G(q)$. Notice that, if we put $E=0$, one pole is canceled out, and $G(q)$ greatly simplifies
\begin{equation}
   G(q) =\frac{1}{1+(t_1/t_2)q}\left(%
\begin{array}{c}
  1 \\
  0 \\
\end{array}\right),
\end{equation}
and the edge state exists if
\begin{equation}
  |t_2/t_1|>1.
\end{equation}

%%%%%%%%%%%%%%%%%%%%%%%%%%%%%%%%%%%%%%%%%%%%%%%%%%%%%%%%%%%%%%%%%%%%%%%%%%%%%
\section{Energy of the edge states in the generalized Shockley model} \label{sec:appendixB}
%%%%%%%%%%%%%%%%%%%%%%%%%%%%%%%%%%%%%%%%%%%%%%%%%%%%%%%%%%%%%%%%%%%%%%%%%%%%%
In this section, we use the generating function method to prove that an edge eigenstate for Hamiltonian~(\ref{BlochHam4}) can exist only for the eigenenergy $E=0$. Like in the previous section, Hamiltonian~(\ref{BlochHam4}) can be given in a recursive form Eq.~(\ref{recur}) with the following $U$ and $V$
\begin{equation}
   V = \left(%
\begin{array}{cc}
  0 & t_2^\ast \\
  t_3 & 0 \\
\end{array}%
\right),\,\,\,\, U = \left(%
\begin{array}{cc}
  0 & t_1^\ast \\
  t_1 & 0 \\
\end{array}%
\right).
\end{equation}
Using Eq.~(\ref{gen}) we obtain the generating function
\begin{equation}
  G(q) = \frac{N(q)}{D(q)}, \label{Gqq}
\end{equation}
where the numerator
\begin{equation}
 N(q)= \left(%
\begin{array}{cc}
 E q & \beta(q) \\
 \alpha(q)  & E q \\
\end{array}%
 \right) \left(%
\begin{array}{c}
  \psi_1 \\
  \psi_2 \\
\end{array}%
\right) \label{Nq}
\end{equation}
and denominator
\begin{eqnarray}
 &&D(q) = \alpha(q)\,\beta(q)-E^2q^2
\end{eqnarray}
are defined through the polynomials
\begin{eqnarray}
 &&\alpha(q) = t_3q^2+t_1q+t_2, \label{alq}\\
 &&\beta(q) = t_2^\ast q^2+t_1^\ast q+t_3^\ast.\label{beq}
\end{eqnarray}
In  Eq.~(\ref{Nq}), the following notation is used for brevity
\begin{equation}
\left(%
\begin{array}{c}
  \psi_1 \\
  \psi_2 \\
\end{array}%
\right) = V^\dag
\left(%
\begin{array}{c}
  \psi_a(1) \\
  \psi_b(1) \\
\end{array}%
\right).
\end{equation}
According to Proposition~1, the poles of Eq.~(\ref{Gqq}) determine whether $G(q)$ corresponds to an edge state. The potential poles of $G(q)$ are given by zeros of the quartic polynomial $D(q)$ in the denominator. Thus, let us find the structure of zeros of $D(q)$. Suppose, $q_1$ is a solution of the quartic equation $D(q_1)=0$. Then, since $\left[D(1/q^\ast)\right]^\ast=D(q)/q^2$, $1/q_1^\ast$ is also a solution of the quartic equation $D(1/q_1^\ast)=0$. So, in the most general case, the polynomial $D(q)$ has zeros $q_1$ and $q_2$, as well as $1/q_1^\ast$ and $1/q_2^\ast$. Thus, according to Proposition~1, the only way to build an edge state is to have the smallest poles $|q_1|<1$ and $|q_2|<1$ canceled out by the zeros of the numerator $N(q)$. So, both components of the vector
\begin{eqnarray}
N(q) = \left(%
\begin{array}{c}
  t_2^\ast\psi_2 q^2+[t_1^\ast\psi_2+E\psi_1] q+t_3^\ast\psi_2\\
  t_3\psi_1q^2+[t_1\psi_1+E\psi_2]q+t_2\psi_1
\end{array}
\right)%
\label{eqw}
\end{eqnarray}
must be proportional to $(q-q_1)(q-q_2)$ and, thus, be linearly dependent. Hence, the coefficients in front of the terms $q^2$ and $q^0$ should also be  linearly dependent and so
\begin{equation}
  \psi_1\psi_2(|t_2|^2-|t_3|^2)=0.
\end{equation}
If $|t_2|\neq|t_3|$, then  $\psi_1\psi_2=0$, so the substitution of $\psi_{1}=0$ and $\psi_{2}\neq 0$ (or vice versa) in Eq.~(\ref{eqw}) and the requirement, that both components of $N(q)$ are proportional, lead to $E=0$. The case $|t_2|=|t_3|$ is trivial, because Vieta's formulas for Eq.~(\ref{eqw}) require that $|q_1q_2|=|t_2/t_3|=1$, which contradicts to the initial assumption that $|q_1|<1$ and $|q_2|<1$. Thus, we have proved that the edge states of Hamiltonian~(\ref{BlochHam4}) can only exist for $E=0$, and  there are no other edge states.

%%%%%%%%%%%%%%%%%%%%%%%%%%%%%%%%%%%%%%%%%%%%%%%%%%%%%%%%%%%%%%%%%%%%%%%%%%%%%
\section{Comparison to the model by Mong and Shivamoggi~\cite{Mong-2011}.} \label{sec:appendixC}
%%%%%%%%%%%%%%%%%%%%%%%%%%%%%%%%%%%%%%%%%%%%%%%%%%%%%%%%%%%%%%%%%%%%%%%%%%%%%
Mong and  Shivamoggi~\cite{Mong-2011} considered the tight-binding model with the Hamiltonian
\begin{eqnarray}
&  \bm H = \sum_z \Psi^\dag(z)\left[U\Psi(z)+V\Psi(z-1)+V^\dag\Psi(z+1)\right], \nonumber \\
 \label{HamAp}
\end{eqnarray}
Here, $U$ and $V$ represent the intra-cell and inter-cell $2\times 2$ matrices
\begin{eqnarray}
  U =\bm\tau\bm b^0,\,\,  V =\bm\tau\bm b, \label{scal}
\end{eqnarray}
where $\bm\tau=(\tau_x,\tau_y,\tau_z)$ are the Pauli matrices in the $AB$ sublattice space,   $\bm b_0$ is a real vector, and $\bm b$ a complex vector. Our Hamiltonian~(\ref{BlochHam4}) can also be written in the form of Eqs.~(\ref{HamAp}) and~(\ref{scal}) with
\begin{eqnarray}
  && \bm b^0 = ({\rm Re}\,t_1,\,{\rm Im}\,\, t_1,h),\\
  && \bm b = (t_2^\ast+t_3,\,i(t_2^\ast-t_3),\,0)/2. \label{ourmodel}
\end{eqnarray}
Notice that $b_z=0$ in Eq.~(\ref{ourmodel}). It can shown that the general Hamiltonian~(\ref{HamAp}) can be always transformed to a form with $b_z=0$. Indeed, the unitary transformations $e^{-i\tau_j\phi}$ generated by the Pauli matrices $\tau_j$ rotate the basis for the vector $\bm b$ in the bilinear form $\bm\tau\bm b$. For an arbitrary complex vector $\bm b$, it is always possible to select the axis $z$ to be orthogonal to both $\bm b$ and $\bm b^\ast$, e.g. along $i(\bm b\times\bm b^\ast)$. So, there always exists a basis where $b_z=0$, and the generalized Shockley model discussed in Sec.~\ref{sec:generalizedShockley} is equivalent to the model studied in Ref.~\cite{Mong-2011}.

In the Fourier representation, Eq.~(\ref{HamAp}) has the form  of Eq.~(\ref{TauRepresent}) with the vector $\bm d = \bm b_0+\bm b e^{-ik}+\bm b^\ast e^{ik}$. When $k$ changes from $0$ to $2\pi$, the vector $\bm d(k)$ stays in the plane spanned by the vectors $(\bm b,\bm b^\ast)$ and offset from the origin by the vector $\bm b_0$.

%%%%%%%%%%%%%%%%%%%%%%%%%%%%%%%%%%%%%%%%%%%%%%%%%%%%%%%%%%%%%%%%%%%%%%%%%%%%%

%%%%%%%%%%%%%%%%%%%%%%%%%%%%%%%%%%%%%%%%%%%%%%%%%%%%%%%%%%%%%%%%%%%%%%%%%%%%%

%%%%%%%%%%%%%%%%%%%%%%%%%%%%%%%%%%%%%%%%%%%%%%%%%%%%%%%%%%%%%%%%%%%%%%%%%%%%%

\begin{thebibliography}{99}
\bibitem{kpNote} Note that this notation is different from the  $(\bm k\cdot\bm p)$ method in the literature.
\bibitem{kNote} For the diamond model, the variable $k$ is the momentum along the translation vector $\bm a_4-\bm a_3$.
\end{thebibliography}
\end{document}